 \documentclass[draft]{agujournal2019}

 \usepackage{url} 
 \usepackage{lineno}
 
\usepackage{soul}
\nolinenumbers
\usepackage{lipsum}
\usepackage{mathtools}
\usepackage{enumerate}

\usepackage{multirow}
\usepackage{longtable}
\usepackage{graphicx}
\usepackage{threeparttable}

 \usepackage{amsmath,amsfonts,amssymb,bm}
 \usepackage{mathptmx}
 \usepackage{newtxtext}
 \usepackage{newtxmath}

 \usepackage{ragged2e}

\usepackage{threeparttable}
\usepackage{float}

\usepackage{extarrows}

  \draftfalse

\journalname{Journal of Advances in Modeling Earth Systems (JAMES)}

\begin{document}

%
%

\title{Global Estimation of Subsurface Eddy Kinetic Energy of Mesoscale Eddies Using a Multiple-input Residual Neural Network
}

%
%




 \authors{Chenyue Xie\affil{1} and An-Kang Gao\affil{1} and Xiyun Lu\affil{1}}
 

\affiliation{1}{Department of Modern Mechanics, University of Science and Technology of China, Hefei, China}




\correspondingauthor{Chenyue Xie}{cyxie@ustc.edu.cn}




\begin{keypoints}
\item Multiple-input neural network models are developed to estimate subsurface EKE using surface and sparse subsurface climatological variables

\item The proposed MI-ResNet model outperforms other data-driven and physical approaches in reconstructing the vertical structure of EKE

\item The locally trained MI-ResNet model can be effectively generalized globally and is applicable to observational data with transfer learning

\end{keypoints}

%
%

%
%


\begin{abstract}
\justifying

Oceanic eddy kinetic energy (EKE) is a key quantity for measuring the intensity of mesoscale eddies and for parameterizing eddy effects in ocean climate models. Three decades of satellite altimetry observations allow a global assessment of sea surface information. However, the subsurface EKE with spatial filter has not been systematically studied due to the sparseness of subsurface observational data. The subsurface EKE can be inferred both theoretically and numerically from sea surface observations but is limited by the issue of decreasing correlation with sea surface variables as depth increases. In this work, inspired by the Taylor-series expansion of subsurface EKE, a multiple-input neural network approach is proposed to reconstruct the subsurface monthly mean EKE from sea surface variables and subsurface climatological variables (e.g., horizontal filtered velocity gradients). Four neural networks are trained on a high-resolution global ocean reanalysis dataset, namely, surface-input fully connected neural network model (FCNN), surface-input Residual neural network model (ResNet), multiple-input fully connected neural network model (MI-FCNN), and multiple-input residual neural network model (MI-ResNet). The proposed MI-FCNN and MI-ResNet models integrate the surface input variables and the vertical profiles of subsurface variables. The MI-ResNet model outperforms the FCNN, ResNet, and MI-FCNN models, and traditional physics-based models in both regional and global reconstruction of subsurface EKE in the upper 2000 m. In addition, the MI-ResNet model performs well for both regional and global observational data based on transfer learning. These findings reveal the potential of the MI-ResNet model for efficient and accurate reconstruction of subsurface oceanic variables.

\end{abstract}

\section*{Plain Language Summary}
\justifying

Mesoscale eddies, which are loosely defined as swirling water masses of tens to hundreds of kilometers wide, are crucial in shaping ocean circulation and influencing global climate dynamics. Eddy kinetic energy (EKE) is a key indicator of mesoscale eddy strength and is essential for building global ocean models and for various applications in oceanography and climate studies. The surface EKE can be calculated from satellite data that measures the sea surface height, but it is much harder to measure EKE below the surface because there are not enough subsurface observations. This work explores machine learning methods to predict subsurface EKE by utilizing sea surface variables and sparse vertical profiles of variables from reanalysis and observational datasets. We develop a multiple-input residual neural network model that can accurately reconstruct global subsurface EKE. This work not only improves the prediction performance of subsurface processes but also contributes to broader applications in climate modeling.

%
%

%


%
%
%
%

\clearpage
\section{Introduction}
\label{Sec: Introduction}
\justifying

Mesoscale eddies with horizontal scales of tens to hundreds of kilometers are ubiquitous in the ocean, covering over 40\% of the global sea surface and containing approximately 90\% of the ocean's total kinetic energy~\cite{Klein-2009,Ferrari-2010}. These eddies are important for the ocean's energy cycle and influence the mean state of the ocean. They transport heat and carbon while mediating energy transfer in mixing process~\cite{Jayne-2002, Chelton-2011, Marshall-2017, Amores-2017}. A key metric for assessing the strength and dynamics of mesoscale eddies is the eddy kinetic energy (EKE), which is an important variable in ocean models and climate studies. For example, it has been parameterized to inform Gent and McWilliams (GM) parameterization and backscatter coefficients in recent years~\cite{Gent-1990-mixing,Cessi-2008,Eden-2008,Marshall-2010, Jansen-2019,Bachman-2019,Juricke-2019,Juricke-2020,Yankovsky-2024}, but reconstructing the vertical structure of EKE in ocean models remains difficult. Satellite observations have revolutionized our understanding of the surface EKE using advanced satellite technologies such as the Surface Water and Ocean Topography (SWOT) and the Guanlan mission~\cite{Reynolds-2007, Fu-2014, Banzon-2016, Wang-2018, Morrow-2019, Chen-2019, Huang-2020, Fu-2024}. Meanwhile, sparse Argo floats can offer in-situ measurements of ocean subsurface temperature and salinity~\cite{Fabienne-2016, Nicolas-2023}. However, estimating the EKE in the subsurface ocean from satellite and sparse Argo measurements is important but also challenging.

To address these challenges, recent studies have shown that the combination of barotropic and baroclinic modes can effectively capture the vertical structure of eddy velocity using sea surface information~\cite{Wunsch-1997, Wunsch-1999, Wortham-2014, Stanley-2020, Ni-2023}. In addition, the ``surface'' modes characterized by zero flow at the ocean bottom have been proposed to describe the vertical structure of EKE~\cite{Wortham-2014, Lama-2016, Groeskamp-2020, LaCasce-2017, LaCasce-2020}. \citeA{Groeskamp-2020} utilized the first surface mode to reconstruct subsurface EKE and further estimated subsurface eddy mixing. \citeA{Stanley-2020} approximated the vertical structure of EKE and transfer coefficient in the GM parameterization with the first two surface modes~\cite{Stanley-2020}. Moreover, integrating satellite altimetry with gridded Argo float data has improved estimation of full-depth EKE utilizing the surface modes~\cite{Ni-2023}. Based on the linear Quasi-Geostrophic (QG) Potential Vorticity (PV) equation, the stream function can be decomposed into a surface Quasi-Geostrophic (SQG) solution and an interior solution~\cite{Hoskins-1975, Lapeyre-2006, Ferrari-2010, Wang-2013, Liu-2014, Liu-2017}. 
The EKE of the sea surface and in-depth in the South Atlantic region has been effectively described using the interior and surface QG method~\cite{Miracca-Lage-2022}. In both flat and rough bottom cases, the normal modes of vertical structure remain independent of horizontal scales~\cite{Lapeyre-2009, Hausmann-2012, Smith-2013, Frenger-2015, Yassin-2022}. Recently, \citeA{Zhang-2024} introduced the SQG-based scale-aware method for estimating the vertical profile of EKE.

In addition to the physics-based methods, machine learning techniques are increasingly being applied to reconstruct various oceanographic variables~\cite{Chapman-2017, Chen-2018, Bolton-2019, Zanna-2020, Guillaumin-2021, Foster-2021, George-2021, Manucharyan-2021, Partee-2021, Pauthenet-2022, Zhuyc-2022, Guan-2022, Tian-2022, Solodoch-2023, Zhu-2023, He-2024, Meng-2024, Mengzl-2024}. The subsurface temperature and velocity fields can be reconstructed with the self-organizing map method using sea surface data~\cite{Chapman-2017, Chen-2018}. \citeA{Bolton-2019} predicted the subsurface stream function using a convolutional neural network (CNN) from surface stream function trained on QG data. Eddy heat fluxes can be derived from sea surface height (SSH) for QG baroclinic turbulence using CNN~\cite{George-2021}. Residual CNNs have proven effective in accurately estimating surface and subsurface stream function from SSH in a two-layer QG model of baroclinic ocean turbulence~\cite{Manucharyan-2021}. The subsurface vertical velocity can be reconstructed with machine learning models based on sea surface variables~\cite{Zhu-2023, He-2024}. The Antarctic Circumpolar Current transport and meridional overturning circulation strength can be reproduced by the CNN model from the SSH and ocean bottom pressure in an idealized channel model~\cite{Meng-2024}.

Although the vertical structure of EKE based on Reynolds averaging has been widely studied~\cite{Lama-2016,Groeskamp-2020,Ni-2023,Torres-2023,Zhang-2024,Yankovsky-2024}, this approach is insufficient in addressing spatial cross-scale energy transfer and developing spatial scale-aware models. In contrast, spatial filtering has been widely employed to investigate the relationship between large and small spatial scales in turbulence~\cite{Smagorinsky-1963, Leonard-1975, Germano-1992, Meneveau-2000}. Recently, it has been applied to study the multiple-scale dynamics in the ocean~\cite{Aluie-2018, Aluie-2019, Rai-2021}. The basic spatial filter used in physical oceanography is the top-hat spatial filter~\cite{Aluie-2018, Grooms-2021, Guillaumin-2021, Uchida-2022, Storer-2022, Ross-2023, Buzzicotti-2023, Contreras-2023, Khani-2023, Yao-2023, Jakhar-2024}. \citeA{Contreras-2023} described the spatiotemporal distribution of turbulent cascades in the Gluf Stream with top-hat spatial filter. \citeA{Khani-2023} developed a new subgrid-scale model to parameterize subgrid mesoscale eddy transports and momentum fluxes based on the spatial filtering approach. The vertical structures of mesoscale eddies in the Kuroshio-Oyashio Extension region are classified with the hierarchical ascending classification~\cite{Yao-2023}. However, the vertical structure of global EKE using spatial filtering remains underexplored, mainly due to the sparse subsurface observational data.

In this study, to accurately reconstruct the subsurface EKE with spatial filtering, we propose a multiple-input neural network framework to reconstruct the vertical structure of EKE using sea surface variables and sparse subsurface climatological variables, based on the Taylor-series expansion of EKE. An additional surface-input neural network framework that only considers sea surface variables is trained for comparison. The performance of these models is evaluated outside the tropics ($[60^\circ\mathrm{S},~10^\circ\mathrm{S}]\ \& \ [10^\circ\mathrm{N},~60^\circ\mathrm{N}]$) on reanalysis and observational data~\cite{Reynolds-2007, Amante-2009, Becker-2009, Banzon-2016, Fabienne-2016, Huang-2020, Jean-Michel-2021, Nicolas-2023}.

The rest of this paper is organized as follows. \S\ref{Sec: Methodology} provides an overview of the reanalysis and observational data, gives the definitions of spatial filter and EKE, \S\ref{Sec: Theoretical and neural network models} describes theoretical and neural network models for reconstructing the vertical structure of EKE. In \S\ref{Sec: Results}, we first evaluate the performance of the aforementioned models using reanalysis data and identify the best model, which is then transferred to the observational data. Finally, \S\ref{Sec: Discussion and conclusion} presents a discussion and conclusion.

\section{Methodology}\label{Sec: Methodology}

\subsection{Eddy-resolving reanalysis and gridded observational datasets}\label{Sec: Datasets}

In this section, we present the datasets used for systematically assessing the performance of different models in predicting global subsurface EKE: a global eddy-resolving reanalysis and gridded observational data. 

As shown in Table \ref{Table: Product}, the NEMO eddy-resolving reanalysis model data (GLORYS12V1, hereafter GLORYS) provides all the essential oceanic variables such as SSH, SST, temperature, salinity, and velocities with a horizontal resolution of 1/12$^\circ$ and 50 vertical levels~\cite{Madec2016, Jean-Michel-2021, Storer-2022, Storer-2023, Buzzicotti-2023, Wangx-2024, Martin-2024}, which assimilates observational data through a reduced-order Kalman filter. The geostrophic velocity and its spatiotemporal spectrum calculated from GLORYS data are highly consistent with the results derived from AVISO satellite data~\cite{Storer-2022, Buzzicotti-2023}. 
In addition, this research uses a gridded bathymetric data with a horizontal resolution of 1/12$^\circ$, which includes ETOPO1 for the deep ocean and GEBCO8 for the coast and continental shelf~\cite{Amante-2009, Becker-2009, Jean-Michel-2021}. The reanalysis data used in this study covers 20 years from January 2001 to December 2020.

\begin{table}[tbh!] 
\scriptsize
\centering
\renewcommand\arraystretch{1}
\caption{List of datasets, oceanic variables, horizontal and vertical resolutions, vertical coverage, temporal frequency, and time periods used in the eddy-resolving reanalysis and gridded observational data.}
\resizebox{1\textwidth}{!}{
\begin{tabular}{lcccccc}
\hline
Datasets &Variable &Description &Horizontal resolution &Vertical coverage and resolution &Temporal frequency &Time period\\
\hline
GLORYS  &$\theta$ &Temperature &1/12$^\circ$ &50 levels (0.494-5727.917 m) & Daily mean     &Jan. 2001 - Dec. 2020\\
            &S &Salinity    &1/12$^\circ$ &50 levels (0.494-5727.917 m)  & Daily mean     &Jan. 2001 - Dec. 2020\\
            &SSH &Sea surface height above geoid &1/12$^\circ$ &Sea surface  & Daily mean     &Jan. 2001 - Dec. 2020\\
            &$u$ &Eastward velocity &1/12$^\circ$ &50 levels (0.494-5727.917 m)  & Daily mean     &Jan. 2001 - Dec. 2020\\
            &$v$ &Northward velocity &1/12$^\circ$ &50 levels (0.494-5727.917 m)  & Daily mean     &Jan. 2001 - Dec. 2020\\
\hline
ETOPO1+GEBCO8  &H &Bathymetry &1/12$^\circ$ &Sea floor depth below geoid   & --   &--\\    
\hline
CMEMS       &SSH   &Sea surface height above geoid   &1/4$^\circ$  &Sea surface  & Daily mean     &Jan. 2002 - Dec. 2020\\
\hline
OISSTv.2.1  &SST   &Sea surface temperature  &1/4$^\circ$ &Sea surface  & Daily mean     &Jan. 2002 - Dec. 2020\\
\hline
ISAS20\_ARGO  &$\theta$ &Temperature &1/2$^\circ$ &187 levels (0-5500 m)   & Monthly mean   &Jan. 2002 - Dec. 2020\\
              &S        &Salinity              &1/2$^\circ$ &187 levels (0-5500 m)   & Monthly mean   &Jan. 2002 - Dec. 2020\\        
\hline
\end{tabular}
}
\label{Table: Product}
\end{table}


Satellite altimetry and gridded Argo float observations are also presented in Table \ref{Table: Product}. The Ssalto/Ducas daily absolute dynamic topography (ADT) from 2002 to 2020 published by the Copernicus Marine and Environment Monitoring Services, has a horizontal resolution of 1/4$^\circ$. The SST is provided by the NOAA Optimum Interpolation Sea Surface Temperature (OISST) dataset~\cite{Reynolds-2007, Banzon-2016, Huang-2020}, which has the same horizontal resolution as the altimeter observations and covers the entire altimeter record period. The In Situ Analysis System (ISAS20) dataset provides monthly mean temperature and salinity, covering 187 standard depth levels from sea surface to 5500 m depth with a horizontal resolution of 1/2$^\circ$~\cite{Fabienne-2016, Nicolas-2023}. Monthly mean in situ density can be derived from monthly mean sea water temperature and practical salinity using the density formula proposed by \citeA{Gill-1982} at zero pressure with the Gibbs-SeaWater Oceanographic Toolbox. To maintain the consistency of the horizontal resolution of the observational data, the monthly mean density is interpolated to a horizontal resolution of  1/4$^\circ$ to match the horizontal resolution of the SSH and SST~\cite{Wangx-2024}. In this research, we focus on the global ocean beyond 10$^\circ$ latitude from the equator, which allows the estimation of geostrophic velocity from the SSH using geostrophic balance~\cite{Vallis-2017}.

The geostrophic approximation at the sea surface assumes that the Coriolis force and the pressure gradient force are in balance, thus establishing a statistical equilibrium. From the geostrophic balance theory, the surface geostrophic velocities ($u_{g}^{s}$ and $v_{g}^{s}$, where $g$ and $s$ are abbreviations of geostrophic and sea surface, respectively) can be derived from the SSH relative to the geoid as follow:
\begin{equation}
    u_{g}^{s}=-\frac{g}{f}\frac{\partial \mathrm{SSH}}{\partial y},
    \label{Eqn: Ug}          
\end{equation}
\begin{equation}
    v_{g}^{s}=\frac{g}{f}\frac{\partial \mathrm{SSH}}{\partial x},
    \label{Eqn: Vg}          
\end{equation}
where $f=2\Omega \mathrm{sin}\varphi$ is the local Coriolis frequency ($\Omega$ and $\varphi$ are the rotation rate of the earth and the latitude, respectively), $g$ denotes gravitational acceleration. Furthermore, using the surface geostrophic velocities as references, the vertical profiles of velocities can be derived from the thermal wind relation, which combines the principles of hydrostatic and geostrophic balances~\cite{Vallis-2017}. 

\subsection{The low-pass spatial filter and EKE}\label{Sec: The low-pass spatial filter and EKE}

In this section, We describe the concept of low-pass spatial filter and provide the definition of EKE. Additionally, the Taylor-series expansion of EKE is derived to explore its dependence on various oceanic variables~\cite{Aluie-2018, Grooms-2021, Guillaumin-2021, Uchida-2022, Storer-2022, Loose-2023, Storer-2023, Ross-2023, Contreras-2023, Sunyq-2023, Jakhar-2024}. By using a low-pass spatial filter, such as top-hat spatial filter, we can decompose the oceanic variables into large-scale structures and small-scale fluctuations and effectively capture the interactions among different spatial scales~\cite{Aluie-2018, Srinivasan-2019, Schubert-2020, Stanley-2020, Grooms-2021, Guillaumin-2021, Uchida-2022, Luyy-2022, xiejw-2023}. In this study, the top-hat spatial filter is implemented by convolution of the oceanic variable $\mathrm{F}$:
\begin{equation}
    \overline{\mathrm{F}}(\mathbf{x})=\int \mathrm{C_{L}}(\mathbf{x},\mathbf{x}^*)\mathrm{F}(\mathbf{x}^*)\mathrm{d}\mathbf{x}^*.
    \label{Eqn: Fx}    
\end{equation}
Here, $\overline{\mathrm{F}}$ is the filtered oceanic variable, $\mathbf{x}=(x,y)$ represents the horizontal coordinate, $\mathbf{x}^*$ serves as a dummy integration variable, and the integration is performed over the horizontal domain. The fluctuation, denoted by $\mathrm{F}'$, is defined as the deviation of the oceanic variable $\overline{\mathrm{F}}$ from the filtered field $\overline{\mathrm{F}}$, $\mathrm{F}'=\mathrm{F}-\overline{\mathrm{F}}$. 

The top-hat kernel is defined as follows~\cite{Stanley-2020,Guillaumin-2021,Uchida-2022,Luyy-2022,xiejw-2023}:
\begin{equation}
\mathrm{C_{L}}(\mathbf{x},\mathbf{x}^*)=\left\{\begin{array}{cc} 
\frac{\mathrm{A^*}}{\Delta^2} & \mathrm{if}\ |\mathbf{x}-\mathbf{x}^*| \leq \Delta/2, \\ 
0 & \mathrm{otherwise}
\end{array}
\right\} \mathrm{I_A}(\mathbf{x}^*),
    \label{Eqn: CL}    
\end{equation}
where $\Delta$ is the filter width, $\mathrm{A}^*$ denotes the grid size, $\mathrm{I_A}(\mathbf{x}^*)$ is a mask function that is zero over land and one over the ocean. It treats the land as zero-velocity water, satisfying no-flow boundary conditions~\cite{Aluie-2019, Storer-2022, Buzzicotti-2023}. Although a `fixed' filter kernel is used at all locations, Only the water area is considered in the denominator when computing the average area. The top-hat filter can be used to parameterize mesoscale eddies and avoid the assumptions of homogeneity and isotropy~\cite{Stanley-2020, Grooms-2021, Uchida-2022, Luyy-2022, xiejw-2023, Loose-2023}. Consistent with previous studies, we use a filter width $\Delta = 1^\circ$, defined as a fixed spatial filtering factor multiplied by the local grid size. This is different from the method that chooses a fixed length scale (e.g., 100~km)~\cite{Stanley-2020, Guillaumin-2021, Uchida-2022, Luyy-2022, xiejw-2023}. 

Following the top-hat spatial filter described above, we define the EKE per unit horizontal area as follows:
\begin{equation}
    \mathrm{EKE}=\frac{1}{2}(\overline{uu}+\overline{vv}-\overline{u}\,\overline{u}-\overline{v}\,\overline{v}),
    \label{Eqn: EKE}      
\end{equation}
which is the difference between the total filtered kinetic energy ($\overline{\mathrm{E}}=\frac{1}{2}(\overline{uu}+\overline{vv}$)) and the resolved kinetic energy ($\overline{\mathrm{E}}_{r}=\frac{1}{2}(\overline{u}\,\overline{u}+\overline{v}\,\overline{v})$). Meanwhile, $\Delta = 1^\circ$ is consistent with the horizontal resolution used in most climate models~\cite{Eyring-2016}, which can be used for reanalysis and observational data described in Section \ref{Sec: Methodology}.

The complexity of EKE arises from the nonlinear interaction between large scale and small scale across the filter width. When the physical variable has the structure $\overline{ab}-\overline{a}\overline{b}$ on a regular uniform grid, it can be expanded using the Taylor-series expansion as follows ~\cite{Clark-1979, Vreman-1996, Anstey-2017, Khani-2023, Jakhar-2024}:
\begin{equation}
    \overline{ab}-\overline{a}\overline{b}=\alpha\frac{\partial \overline{a}}{\partial x_{k}}\frac{\partial \overline{b}}{\partial x_{k}}+\cdot\cdot\cdot,
    \label{Eqn: Fg}          
\end{equation}
where $\alpha$ is the coefficient and the detailed derivation can be found in the work of \citeA{Jakhar-2024}. The correlation and similarity of EKE with ($\frac{\partial \overline{u}}{\partial x_{k}}\frac{\partial \overline{u}}{\partial x_{k}}+\frac{\partial \overline{v}}{\partial x_{k}}\frac{\partial \overline{v}}{\partial x_{k}}$) has been widely examined in turbulence and the ocean~\cite{Clark-1979, Vreman-1996, Anstey-2017, Khani-2023, Jakhar-2024}. Equation~(\ref{Eqn: Fg}) and recent machine learning studies on mesoscale eddy parameterization suggest that filtered velocity gradients can play an important role in reconstructing EKE.

\begin{figure}[tbh!]
 \begin{center}
   \includegraphics[width=1\linewidth]{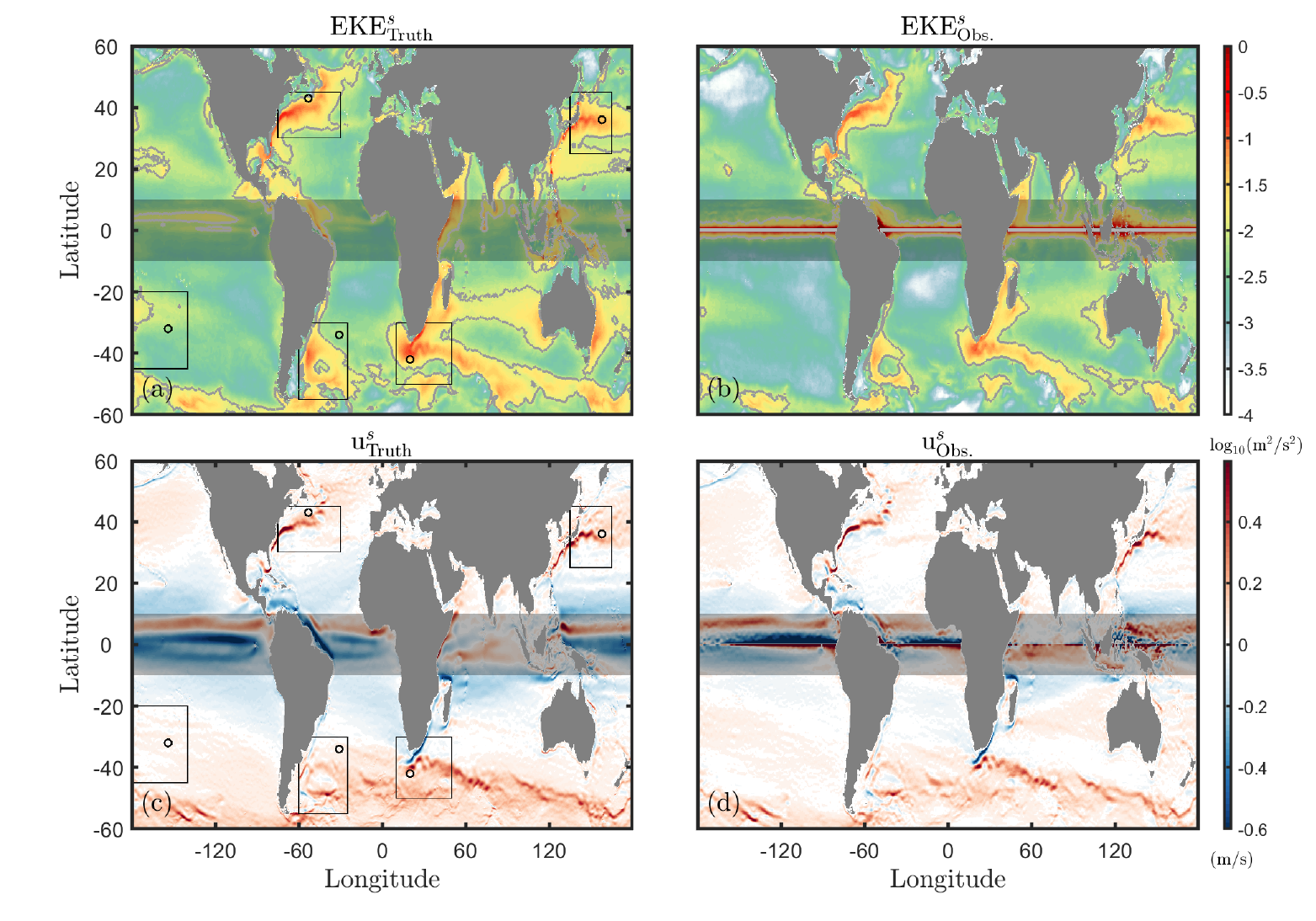}  
   \caption{The multiple-year averaged surface EKE and meridional velocity ($u$) as functions of longitude and latitude using the eddy-resolving GLORYS reanalysis and altimetric data for the period 2001-2020: (a) $\mathrm{EKE}_\mathrm{Truth}^{s}$, (b) $\mathrm{EKE}_\mathrm{Obs.}^{s}$, (c) $\mathrm{u}_\mathrm{Truth}^{s}$, (d) $\mathrm{u}_\mathrm{Obs.}^{s}$. The latitude region between $10^\circ\mathrm{S}$ and $10^\circ\mathrm{N}$ is shaded in grey. Gray dashed lines in (a) and (b) denote the isopleths of EKE with $0.01~\mathrm{m}^2/\mathrm{s}^2$. The five representative regions are chosen for training the neural network models.}
   \label{Fig: EKE_u}
 \end{center}
\end{figure}


EKE with top-hat filter emphasizes nonlinear interactions across spatial scales. In this research, the EKE reproduced from the reanalysis data is used as the ground-truth against which the reconstructed EKE from different models are compared.~\cite{Martin-2024}. In order to examine the consistency of the reanalysis and observational data, the 20-year averaged surface EKE is shown in Figures~\ref{Fig: EKE_u}a--b. $\mathrm{EKE}^{s}_\mathrm{Truth}$ is derived from the sea surface velocities of the reanalysis data, while $\mathrm{EKE}^{s}_\mathrm{Obs.}$ is calculated using sea surface geostrophic velocities derived from satellite observational data. High EKE (greater than 0.01 m$^2/$s$^2$) is mainly located along the ocean currents, such as the Gulf Stream, Kuroshio current, Agulhas current, Brazil-Malvinas current, and within the Southern Ocean~\cite{Beech-2022, Ding-2024}. The high EKE regions reproduced from the reanalysis data and the observational data are consistent. However, the EKE reproduced from the observational data is weaker than that calculated from the reanalysis data, as the horizontal resolutions of the two datasets are 1/4$^\circ$ and 1/12$^\circ$, respectively. The smaller scales are filtered out by the observational data. In addition, Figures~\ref{Fig: EKE_u}c--d show surface velocities calculated from the reanalysis and observational data, respectively. The surface geostrophic velocity $\mathrm{u}_\mathrm{Obs.}^{s}$ is in agreement with $\mathrm{u}_\mathrm{Truth}^{s}$ for regions away from the equator, which is consistent with Equation~(\ref{Eqn: Ug}).

\section{Theoretical and neural network models}\label{Sec: Theoretical and neural network models}

In this section, we outline the approaches used to reconstruct subsurface EKE from sea surface variables and sparse subsurface climatological variables. Two types of models are described in this study: traditional theoretical models and neural network-based models.

\subsection{Traditional models for the vertical structure of EKE}\label{Sec: Theory}

Recent studies have emphasized the importance of reconstructing the vertical structure of EKE based on linear modes~\cite{Groeskamp-2020, Ni-2023, Yankovsky-2024}. The barotropic and first baroclinic modes dominate the ocean's velocity in the extratropical ocean~\cite{Wunsch-1997}. In addition, \citeA{Lama-2016} and \citeA{LaCasce-2017} introduced surface modes with zero flow at the ocean bottom. The linear QGPV equation has been widely used to reconstruct subsurface oceanic variables from satellite altimetry and gridded Argo float observations~\cite{Pedlosky-1987, LaCasce-2012, Vallis-2017}. The QGPV equation is given by:
\begin{equation}
    \frac{\partial}{\partial t}\left[\nabla^2\psi+\frac{\partial}{\partial z}(\frac{f^2}{N^2}\frac{\partial\psi}{\partial z})\right]+\beta\frac{\partial\psi}{\partial x}=0.
    \label{Eqn: LQGPV}          
\end{equation}
Here $\beta$ is the meridional derivative of $f$, $N$ is the Brunt-Vaisala frequency, $\nabla^2=\partial_{x}^2+\partial_{y}^2$, and $\psi(x,y,z,t)$ denotes the geostrophic stream function.

Assuming that there is a wave-like solution in the horizontal space for Equation~(\ref{Eqn: LQGPV}), $\psi(x,y,z,t)=\Phi(z)\hat{\psi}e^{ikx+ily-i\omega t}$, where $k(l)$ is the zonal (meridional) wavenumber. By substituting this expression into Equation~(\ref{Eqn: LQGPV}), we can obtain the eigenvalue problem for the vertical structure, $\Phi(z)$:
\begin{equation}
    \frac{d}{dz}\left(\frac{f^2}{N^2}\frac{d\Phi}{dz}\right)=-\lambda^{2}\Phi,
    \label{Eqn: Phi}         
\end{equation}
where $\lambda=[k^2+l^2+(\beta k)/\omega]^{1/2}$ and $\omega$ is the frequency.

Solving Equation~(\ref{Eqn: Phi}) requires appropriate sea surface and ocean bottom boundary conditions. Typically, flat bottom and rigid-lid conditions are assumed to determine the vertical structure of modes where the buoyancy vanishes:
\begin{equation}
    \frac{d\Phi}{dz}=0,\ z=-H,\ 0.
    \label{Eqn: Phi_boundary_BC}          
\end{equation}
This forms a Strurm-Liouville problem for $\Phi$, yielding a set of discrete vertical modes $\{\Phi_m\}$ with corresponding eigenvalues $\{\lambda_{m}^2\}$. The zeroth eigenmode with $m=0$ represents the barotropic mode, while the higher-order modes with $m>0$ are the baroclinic modes~\cite{LaCasce-2012}.

Considering that topography can significantly affect the bottom flow, the ``surface'' modes with zero bottom flow have different boundary conditions compared to standard baroclinic modes and have the following forms~\cite{Wortham-2014, Lama-2016, Groeskamp-2020, LaCasce-2017, LaCasce-2020}:
\begin{equation}
    \Phi=0,\quad z=-H \quad\mathrm{and}\quad \frac{d\Phi}{dz}=0,\quad z=0.
    \label{Eqn: Phi_boundary_SM}              
\end{equation}
The discrete surface modes can be obtained from Equations~(\ref{Eqn: Phi}) and (\ref{Eqn: Phi_boundary_SM}). The first baroclinic mode (BC1) and first surface mode (SM1) can be computed using inverse power iteration and finite volume approximation. Meanwhile, $N$ in Equation~(\ref{Eqn: Phi}) satisfies the following constraint: $N<10^{-5}~\mathrm{s}^{-1}$ is replaced by $10^{-5}~\mathrm{s}^{-1}$~\cite{Stanley-2020}.

Assuming that EKE has a separable vertical structure $\Phi^{2}$, which can be expressed as~\cite{Groeskamp-2020, Ni-2023, Yankovsky-2024}:
\begin{equation}
    \mathrm{EKE}=\mathrm{EKE}^{s}\Phi^2(z),
    \label{Eqn: EKE_theory}      
\end{equation}
where $\Phi(z)$ can be represented by the first baroclinic mode ($\Phi_\mathrm{BC1}$) and first surface mode ($\Phi_\mathrm{SM1}$). The sea surface $\mathrm{EKE}^{s}$ can be reproduced from the reanalysis and observational data, and is also affected by the horizontal resolution of data.

\subsection{Neural network-based models for the vertical structure of EKE}\label{Sec: Neural network problem}

Compared to the theoretical models outlined in the previous section that reconstruct the vertical structure of EKE using Equation~(\ref{Eqn: EKE_theory}), the neural network approach provides new insights by effectively capturing the nonlinear relationship between sea surface and sparse subsurface variables and the vertical profile of EKE. In this section, we introduce two neural network frameworks: surface-input neural network and multiple-input neural network. Additionally, we discuss the fully connected neural network (FCNN) and the residual neural network (ResNet) within each framework.

\subsubsection{Surface-input neural network models}\label{Sec: Surface-input neural network models}

As shown in Table \ref{Table: Network comparison}, in order to evaluate the performance of predicting EKE using only sea surface variables, we consider two neural network models based on the sea surface information~\cite{George-2021, Manucharyan-2021, Solodoch-2023}: FCNN and ResNet models.

\begin{table}[ht]
\scriptsize
\centering
\caption{List of the surface input variables, subsurface input variables, residual blocks, dropout, and loss function used in different neural network models.}
\resizebox{1\textwidth}{!}{
\begin{tabular}{lccccc}
\hline
Model & Surface Input Variables & Subsurface Input Variables & Residual Blocks &Dropout &Huber Loss Function\\
\hline
FCNN & $\checkmark$ & $\times$ & $\times$ & $\times$ & $\checkmark$   \\
MI-FCNN & $\checkmark$ & $\checkmark$ & $\times$ & $\times$ & $\checkmark$  \\
ResNet & $\checkmark$ & $\times$ & $\checkmark$ & $\checkmark$ & $\checkmark$  \\
MI-ResNet & $\checkmark$ & $\checkmark$ & $\checkmark$ & $\checkmark$ & $\checkmark$  \\
\hline
\end{tabular}}
\label{Table: Network comparison}
\end{table}

\begin{table}[tbp!]
\scriptsize
\centering
\caption{Raw forms of input and output variables for the neural network models (FCNN, MI-FCNN, ResNet, and MI-ResNet). The models (FCNN and ResNet) used for learning the EKE are allocated with the sea surface input variables of the first line, which exclude the subsurface variables. The models (MI-FCNN and MI-ResNet) used for learning the EKE are allocated with all selected sea surface and subsurface input variables. The reader is referred to \S\ref{Sec: Neural network problem} for the precise definitions of selected variables.}
\resizebox{1\textwidth}{!}{
\begin{tabular}{lccc}
\hline
Models &Surface branch inputs &Subsurface branch inputs  &Outputs\\ \hline
{FCNN, ResNet} &$\mathrm{SSH}$,~$\mathrm{SST}$,~H,~$\beta_t$,~$\overline{u}^{s}_{g}$,~$\overline{v}^{s}_{g}$,~$\frac{\partial \overline{u}^{s}_{g}}{\partial x}$,~$\frac{\partial \overline{u}^{s}_{g}}{\partial y}$,~$\frac{\partial \overline{v}^{s}_{g}}{\partial x}$,~$\frac{\partial \overline{v}^{s}_{g}}{\partial y}$  &\rule{5mm}{0.08mm}  &EKE\\
\hline
{MI-FCNN, MI-ResNet} &$\mathrm{SSH}$,~$\mathrm{SST}$,~H,~$\beta_t$,~$\overline{u}^{s}_{g}$,~$\overline{v}^{s}_{g}$,~$\frac{\partial \overline{u}^{s}_{g}}{\partial x}$,~$\frac{\partial \overline{u}^{s}_{g}}{\partial y}$,~$\frac{\partial \overline{v}^{s}_{g}}{\partial x}$,~$\frac{\partial \overline{v}^{s}_{g}}{\partial y}$ 
&$\overline{\rho}$,~$\overline{u}_{g}$,~$\overline{v}_{g}$,~$\frac{\partial \overline{u}_{g}}{\partial x}$,~$\frac{\partial \overline{u}_{g}}{\partial y}$,~$\frac{\partial \overline{v}_{g}}{\partial x}$,~$\frac{\partial \overline{v}_{g}}{\partial y}$   &EKE\\
\hline
\end{tabular}
}
\label{Table: Network inputs}
\end{table}

FCNNs have been widely used in ocean-related problems, such as parameterizing mesoscale eddies~\cite{Xie-2023a, Xie-2023b}. This study uses an FCNN model with three hidden layers containing 64, 256, and 128 neurons. The activation function used in each hidden layer is the LeakyReLU activation function, defined as follows~\cite{Barth-2020, Manucharyan-2021, Xie-2021, Xie-2023a, Xie-2023b}:
\begin{equation}
\sigma(x)=\left\{
\begin{aligned}
\quad x & , & \mathrm{if}\ x> 0, \\
   0.2x & , & \mathrm{if}\ x\leq0.
\end{aligned}
\right.
  \label{Eqn: Leaky-ReLU}
\end{equation}
A linear activation function is applied to the final output layer. In addition, we demonstrate that increasing the number of neurons in each hidden layer does not significantly affect the results.

The ResNet architecture has been used to infer eddy heat fluxes by~\citeA{George-2021}. In this study, the ResNet model contains three residual blocks with 64, 128, and 64 channels, respectively. Each block contains two convolutional layers with a kernel size of (5,5), and boundary padding is applied to ensure that the output size matches the input size. Each convolutional layer is followed by a batch normalization layer and a dropout layer with a dropout rate of 0.3. After each residual block, a 2D average pooling layer with pooling size (2,2) is applied. Skip connections help maintain gradient flow during training, hence alleviating the vanishing gradient problem. After the residual blocks, two fully connected hidden layers are connected, containing 256 and 128 neurons, respectively. Similar to the FCNN model, all hidden layers in the ResNet architecture are activated by the Leaky-ReLU activation function. The final output layer uses a linear activation function. The weights in the neural networks are regularized using L2 regularization with a regularization parameter of $10^{-5}$.

Next, we detail the surface input variables for the FCNN and ResNet models, which can be obtained through reanalysis product and global observations (i.e., high-frequency and high-resolution satellite data). As shown in Table~\ref{Table: Network inputs}, the inputs include sea surface and topography variables, including SSH, SST, seafloor depth below geoid (H), topographic potential vorticity gradient ($\beta_t=\sqrt{\mathrm{H}_x^2+\mathrm{H}_y^2}$), sea surface filtered geostrophic velocities ($\overline{u}^{s}_{g}$,~$\overline{v}^{s}_{g}$), and sea surface filtered geostrophic velocity gradients ($\frac{\partial \overline{u}^{s}_{g}}{\partial x}$,~$\frac{\partial \overline{u}^{s}_{g}}{\partial y}$,~$\frac{\partial \overline{v}^{s}_{g}}{\partial x}$,~$\frac{\partial \overline{v}^{s}_{g}}{\partial y}$). SSH and SST are included because they are available from altimetry, and both are affected and mediated by ocean dynamics~\cite{Martin-2023, Archambault-2024}. Machine learning models can effectively use them to more accurately reconstruct unobserved oceanic variables~\cite{Liu-2021, George-2021, Martin-2023, Martin-2024}. Considering the complexity of seafloor topography in different latitudes and longitudes, incorporating terrain-related factors H and $\beta_t$ into the inputs can enhance the generalization of the neural networks and emphasize the importance of uneven seafloor~\cite{Lama-2016, LaCasce-2017}. By performing Taylor-series expansion of EKE according to Equation~(\ref{Eqn: Fg}), it can be approximated by a series of filtered velocity gradients~\cite{Clark-1979, Vreman-1996, Anstey-2017, Khani-2023, Jakhar-2024}. Furthermore, subsurface velocity and surface geostrophic velocity can be linked through the thermal wind relation, thus we add surface geostrophic velocities and surface geostrophic velocity gradients to the inputs to increase the prediction performance of the networks. Besides, the reference density and gravitational acceleration used in this study are $\rho_{0}=1025~\mathrm{kg}~\mathrm{m}^{-3}$ and $g=9.81~\mathrm{m}~\mathrm{s}^{-2}$, respectively. 

The horizontal structure of the surface inputs of the FCNN and ResNet models are different: the surface input variables for the FCNN model are taken from the same horizontal coordinate of EKE. In contrast, the ResNet model is highly effective in processing the spatial characteristics of the inputs. The input tensor has a shape of $17 \times 17\times 10$, covering a $4^\circ\times4^\circ$ area around the horizontal coordinate of EKE. The tensor is sampled at a spatial interval of $1/4^\circ$ in latitude and longitude~\cite{Pauthenet-2022, Martin-2023}. The 10 channels represent 10 sea surface variables, as shown in Table~\ref{Table: Network inputs}.

\subsubsection{Multiple-input neural network models}\label{Sec: MI-ResNet}

\begin{figure}[tbh!]
 \begin{center}
   \includegraphics[width=0.9\linewidth]{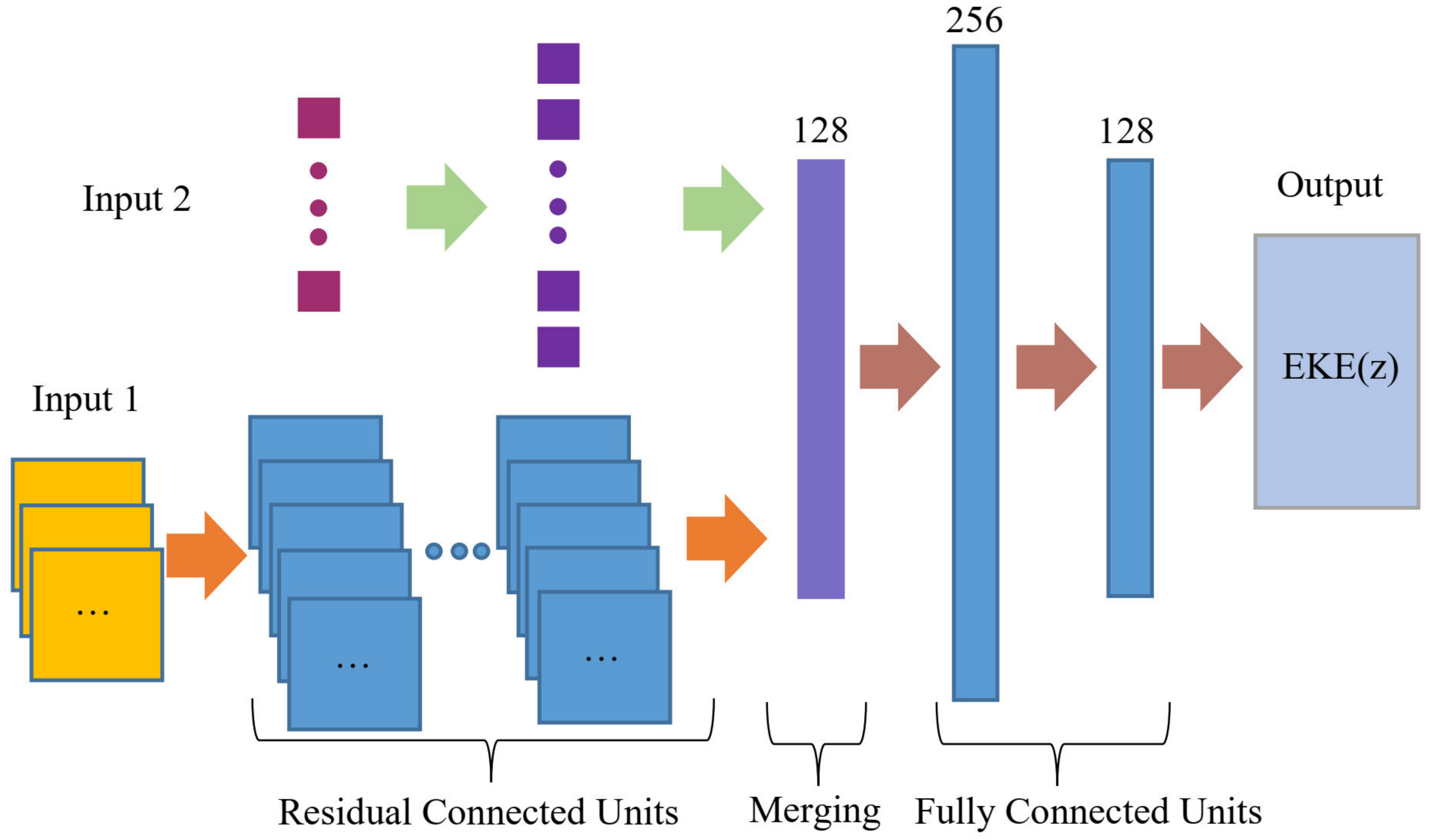}
   \caption{Schematic diagram of the MI-ResNet structure used in this study for estimating subsurface EKE. The first branch of inputs (Input 1) denotes sea surface variables, the second branch of inputs (Input 2) denotes the vertical profiles of variables, and the output of the network is the profile of EKE.}
   \label{Fig: Network}
 \end{center}
\end{figure}

In addition to satellite altimetry observations, the sparse Argo float measurements can also be used to improve the model's performance. In this section, we propose two multiple-input (MI) neural network models to effectively incorporate sparse subsurface information: MI-FCNN and MI-ResNet. Both MI-FCNN and MI-ResNet models integrate two branches~\cite{Fang-2019}: sea surface inputs and the subsurface inputs. The structure of the MI-ResNet model is shown in Figure~\ref{Fig: Network}. 

The MI-FCNN model first processes sea surface and subsurface inputs through separate hidden layers with 64 neurons, respectively. The outputs of these branches are then combined and further processed by the second and third hidden layers containing 256 and 128 neurons. Similar to the MI-FCNN model, the subsurface branch of the MI-ResNet model consists of a hidden layer with 64 neurons. Meanwhile, the surface branch's output is converted into a 64-channel output through 2D global average pooling, then merged with the subsurface branch's 64-neuron output to form a 128-neuron vector. Two hidden layers, containing 256 and 128 neurons, integrate the characteristics from both branches, allowing the MI-ResNet model to enhance the prediction of the vertical structure of EKE. The hidden layers and output layer of the MI-FCNN and MI-ResNet models are activated by LeakyReLU activation function and linear activation function, respectively.

The subsurface branch's inputs are taken from the same horizontal coordinate of EKE. As shown in Table \ref{Table: Network inputs}, the subsurface variables include the vertical profiles of density $\overline{\rho}$, subsurface filtered velocities ($\overline{u}_{g}$,~$\overline{v}_{g}$), and subsurface filtered velocity gradients ($\frac{\partial \overline{u}_{g}}{\partial x}$,~$\frac{\partial \overline{u}_{g}}{\partial y}$,~$\frac{\partial \overline{v}_{g}}{\partial x}$,~$\frac{\partial \overline{v}_{g}}{\partial y}$). These variables provide comprehensive information about the characteristics of subsurface Ocean. For convenience of analysis, the vertical profiles of inputs and EKE are re-interpolated into five vertical layers from surface to 2000~m depth with 500~m intervals. 

A comprehensive comparison of the above four neural network models reveals the following order of model complexity: FCNN$<$MI-FCNN$<$MI-ResNet and FCNN$<$ResNet$<$MI-ResNet. The MI-ResNet model incorporates more sea surface information than the MI-FCNN model and is expected to have better performance in predicting EKE near the surface. Compared with the ResNet model, the MI-ResNet model includes an additional branch for subsurface data, which is expected to improve its performance in predicting deep ocean EKE.

\subsubsection{Training and validation details}\label{Sec: Training details}

The training process is presented in detail after introducing the four neural network models. As shown in Figure~\ref{Fig: EKE_u}, in order to comprehensively capture the global characteristics of EKE across different geographical regions, five representative regions (Gulf Stream, Kuroshio current, Agulhas current, Brazil-Malvinas current, and South Pacific Ocean) are selected for training~\cite{Rieck-2015, Serazin-2018, Juricke-2020, Guillaumin-2021, Meng-2021, Zhu-2022, Feng-2022, Guo-2022, Torres-2023, Gao-2023, Sun-2023}. The Gulf Stream, Kuroshio current, Agulhas current, and Brazil-Malvinas current regions have strong EKE and belong to the western boundary currents and Southern Ocean,  both of which are known for their high EKE. Meanwhile, the EKE in the South Pacific is weak. The five regions account for approximately 14.5\% of the global ocean, spanning from $60.0^\circ\mathrm{S}$ to $10.0^\circ\mathrm{S}$ and from $10.0^\circ\mathrm{N}$ to $60.0^\circ\mathrm{N}$, as shown in Table \ref{Table: Training region}. 

\begin{table}[tbh!]
\small
\centering
\renewcommand\arraystretch{1}
\caption{The five representative regions selected for training the neural network models (FCNN, ResNet, MI-FCNN, and MI-ResNet).}
\resizebox{1.0\textwidth}{!}{
\begin{tabular}{lcccc}
\\[1pt]
\hline
Region &Abbreviation &Latitude range &Longitude range &Surface area/Global ocean (\%)\\
\hline
The Gulf Stream region &GSR &$30^\circ-45^\circ\mathrm{N}$ &$75^\circ-30^\circ\mathrm{W}$ &2.50\% \\
The Kuroshio region &KR &$25^\circ-45^\circ\mathrm{N}$ &$135^\circ-165^\circ\mathrm{E}$ &2.26\% \\
The Agulhas region &AR &$30^\circ-50^\circ\mathrm{S}$ &$10^\circ-50^\circ\mathrm{E}$ &2.77\% \\
The Brazil-Malvinas current region &BMCR &$30^\circ-55^\circ\mathrm{S}$ &$60^\circ-25^\circ\mathrm{W}$ &2.90\% \\
The South Pacific Ocean region &SPOR &$20^\circ-45^\circ\mathrm{S}$ &$180^\circ-140^\circ\mathrm{W}$ &4.05\% \\
\hline
\end{tabular}}
\label{Table: Training region}
\end{table}

As shown in Table~\ref{Table: Network inputs}, the daily mean sea surface filtered velocity gradients are derived from the surface geostrophic velocities with a horizontal resolution of 1/4$^\circ$, which is consistent with the horizontal resolution of Satellite altimetry observations shown in Table~\ref{Table: Product}. The monthly mean subsurface filtered velocity gradients are calculated at 1/2$^\circ$ resolution, the same horizontal resolution as the gridded Argo float observations. In this research, we reconstruct the seasonal variation of EKE by adopting a 1-month rolling mean filter to mitigate the impact of high-frequency variations on the monthly timescale~\cite{Jean-Michel-2021, Hogg-2022, Martinez-Moreno-2022}. The monthly mean EKE of the reanalysis dataset is used as the target function of the neural network models.

To ensure consistency across neural network models, the $z$-score normalization technique is applied to the input and output of the neural networks by subtracting the temporal mean and dividing by the temporal standard deviation~\cite{Archambault-2024}. This method normalizes different oceanic variables to the same scale, eliminating the impact of magnitude differences among inputs and output and ensuring that their means are zero and variances are one. The dataset is divided into two parts: data from 2001 to 2016 is used for training and validating the neural network models, while data from 2017 to 2020 is reserved as an independent test set to evaluate the models on unseen data. A five-fold cross-validation method is used for robust Training. This process requires dividing the training dataset into $k=5$ folds based on temporal order, and the models are trained on $k-1$ folds while retaining one fold for validation. We iterate this process five times~\cite{Martin-2023}. During training across the $k-1$ folds, the data is thoroughly shuffled in both spatial and temporal dimensions to allow for joint training on data from all selected regions simultaneously~\cite{Guillaumin-2021}.

\begin{figure}[tbh!]
 \begin{center}\includegraphics[width=0.9\linewidth]{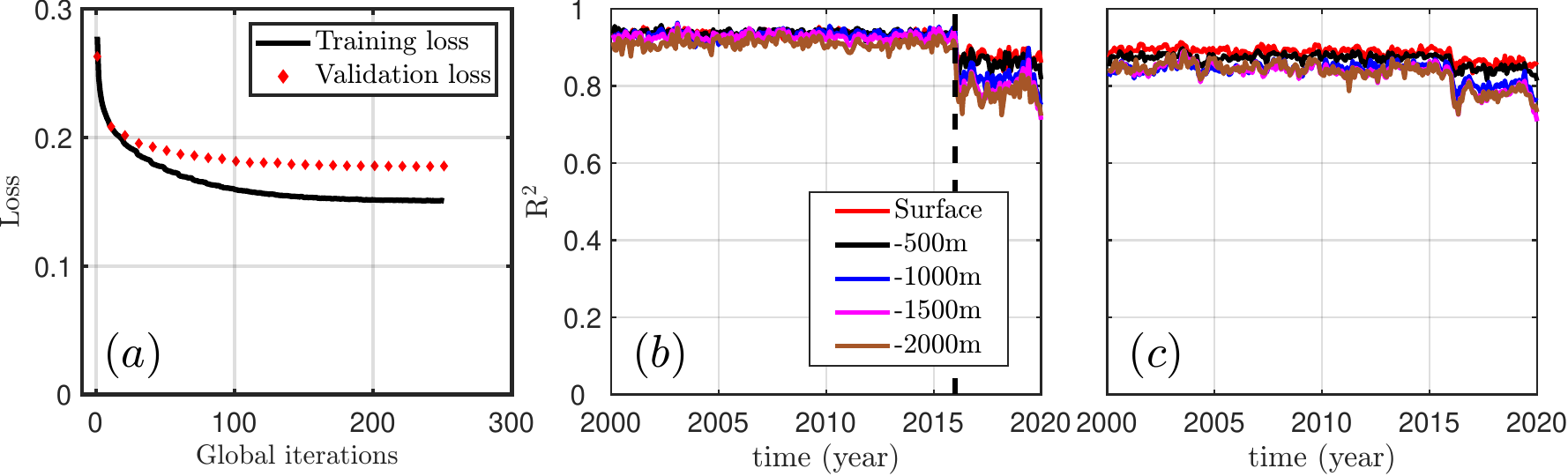}
   \caption{Performance of the MI-ResNet model in predicting EKE. (a) the loss function of the neural network uses the training data from 2001 to 2016, with the first 4/5 of the data for training and the last 1/5 as validation data, (b) the ensemble-averaged time series of $\mathrm{R}^2$ for the training regions, (c) the ensemble-averaged time series of $\mathrm{R}^2$ for the global ocean. The dashed black line in panel (b) divides data into training and test datasets}
   \label{Fig: Loss_function_R2}
 \end{center}
\end{figure}

In this research, the neural networks are trained with the Adam optimizer~\cite{Kingma-2014}. The initial learning rate is set to $1.0\times10^{-3}$, which decays by a factor of 0.8 every 10 epochs, with a minimum threshold of $10^{-5}$ to ensure stable convergence. Early stopping is implemented by monitoring the validation loss function to prevent overfitting. The training process is terminated if the validation loss does not decrease for 20 epochs. 
As shown in Table~\ref{Table: Network inputs}, the Huber loss function is used with the definition~\cite{Huber-1964}:
\begin{equation}
L_{\delta}(a)=\left\{
\begin{aligned}
\quad \frac{1}{2}a^2 & , & \mathrm{for}\ |a|\leq \delta, \\
   \delta\cdot(|a|-\frac{1}{2}\delta) & , & \mathrm{otherwise},
\end{aligned}
\right.
  \label{Eqn: Huber-l}
\end{equation}
where the threshold $\delta=1$. The batch size is 512 to balance the efficiency and stability of training. Figure~\ref{Fig: Loss_function_R2}a shows that the training converges effectively after 200 epochs, and is terminated around 250 epochs.

To evaluate the performance of different models proposed in this study, we use two main statistical metrics: the volume-weighted~\citeA{Nash-1970} coefficient of efficiency $R^{2}$ (also detailed in~\cite{McCuen-2006, Ritter-2013}) and the volume-weighted relative error $E_{r}$, defined as follows~\cite{George-2021}:
\begin{equation}
\begin{cases}
          R^{2}=1-\frac{\langle(\mathrm{EKE}_\mathrm{Truth}-\mathrm{EKE}_\mathrm{Model})^2\rangle}{\langle(\mathrm{EKE}_\mathrm{Truth}-\langle \mathrm{EKE}_\mathrm{Truth}\rangle)^2\rangle},\\
          E_{r}=\frac{\sqrt{\langle(\mathrm{EKE}_\mathrm{Truth}-\mathrm{EKE}_\mathrm{Model})^2\rangle}}{\sqrt{\langle (\mathrm{EKE}_\mathrm{Truth})^2 \rangle}},\\
          \end{cases}\quad          
    \label{Eqn: R2_Er}      
\end{equation}
\noindent
where, $\mathrm{EKE}_\mathrm{Truth}$ is the ``ground truth'' reproduced from the eddy-resolving reanalysis data, and $\mathrm{EKE}_\mathrm{Model}$ represents the EKE reconstructed from different models. $\langle \bullet\rangle=\frac{\iint \bullet ~dxdy}{\iint dxdy}$ denotes a spatial averaging operation over the selected region. The coefficient of efficiency $R^2$ quantifies how well the model predicted EKE matches the ground truth~\cite{Wei-2021, Wei-2022}, with values close to 1 indicating superior performance. The relative error $E_{r}$ measures the prediction errors of the model-predicted EKE relative to the ground truth. These metrics provide a robust assessment of the model's performance in different geographical regions.

\section{Results}\label{Sec: Results}

In this section, we first analyze the performance of different models mentioned in the previous section on reanalysis and observational data in the five training regions shown in Figure~\ref{Fig: EKE_u} and Table~\ref{Table: Training region}. We then further evaluate the performance of the MI-ResNet model over the global ocean. Any spatial locations in the ocean with missing values are excluded from the analysis.

Figures~\ref{Fig: Loss_function_R2}b-c show the temporal evolution of the average $R^2$ of the MI-ResNet model over the training region and global ocean. The temporal-averaged $R^2$ from 2001 to 2016 are greater than 0.91 at different depths for the training region. Meanwhile, for the global ocean, the temporal-averaged $R^2$ exceeds 0.89 and 0.94 at the surface and 2000~m depth, respectively. The lower $R^2$ for the global ocean compared to the training region can be attributed to the fact that the global ocean includes regions outside the training dataset, which is also part of the test dataset. The temporal-averaged $R^2$ from 2017 to 2020 for the training region and global ocean are over 0.85 and 0.77 at the surface and 2000~m depth, respectively. The high $R^2$ calculated from both the training region and global ocean indicate that the MI-ResNet model performs well in both the untrained time range and spatial region, further proving the robustness and generalizability of the MI-ResNet model.

Furthermore, since reliable observational data for training neural network model in real-world applications is usually limited, an effective strategy is to utilize transfer learning to enhance model performance. This involves initially training the neural network model on a closely related reanalysis data with sufficient data, and then fine-tuning it using observational data~\cite{Pan-2010, Yosinski-2014, Foster-2021, Xiao-2023, Archambault-2024, Sun-2024}. In this study, we employ transfer learning with pre-trained weights from the MI-ResNet model, originally trained on eddy-resolving GLORYS reanalysis data. The model is then fine-tuned to estimate EKE using inputs from satellite and gridded Argo observational data~\cite{Pan-2010, George-2021}. The training and validation datasets cover the period from 2002 to 2016, while the test dataset spans from 2017 to 2020. This approach enables us to leverage the ``knowledge'' accumulated in the pre-trained MI-ResNet model and address the challenge of limited observational data.

\subsection{The regions with intense eddies and strong currents}\label{Sec: Intense eddies regions}

\begin{figure}[tbh!]
 \begin{center}
  \includegraphics[width=0.9\linewidth]{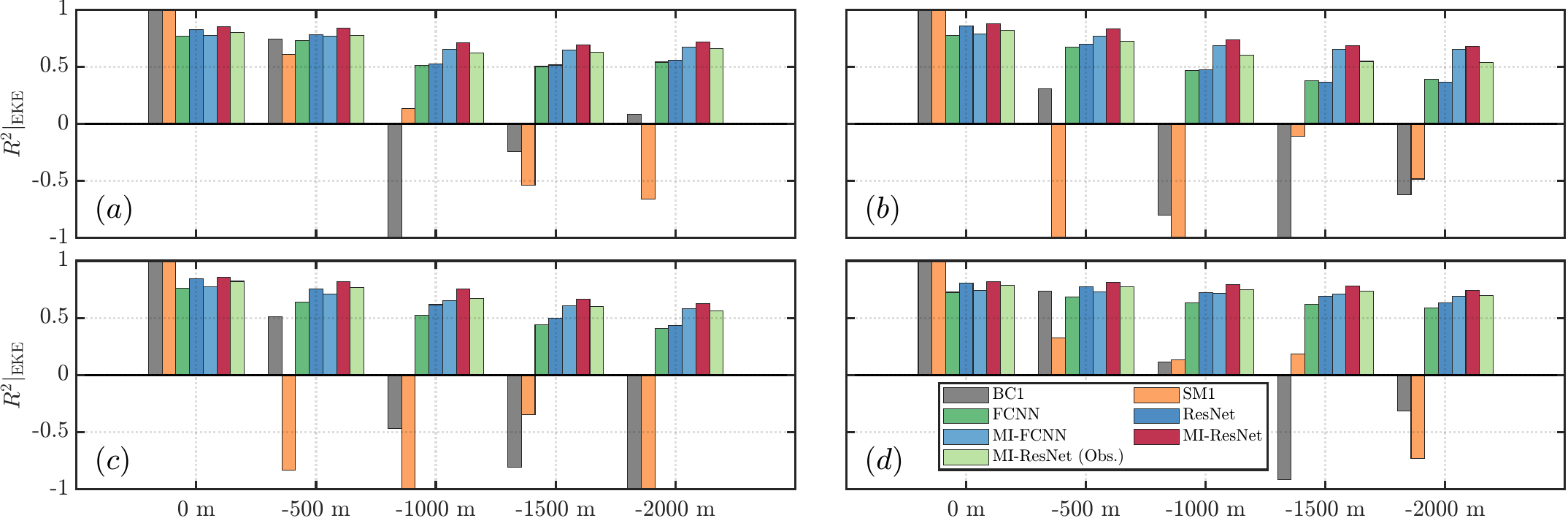}
   \caption{The temporally averaged coefficients of efficiency $R^2$ of $\mathrm{EKE}$ produced by different models (BC1, SM1, FCNN, ResNet, MI-FCNN, and MI-ResNet) for the test dataset (2017-2020): (a) GSR, (b) KR, (c) AR, (d) BMCR.}
   \label{Fig: R2_region_four}
 \end{center}
\end{figure}

\begin{figure}[tbh!]
 \begin{center}
  \includegraphics[width=0.9\linewidth]{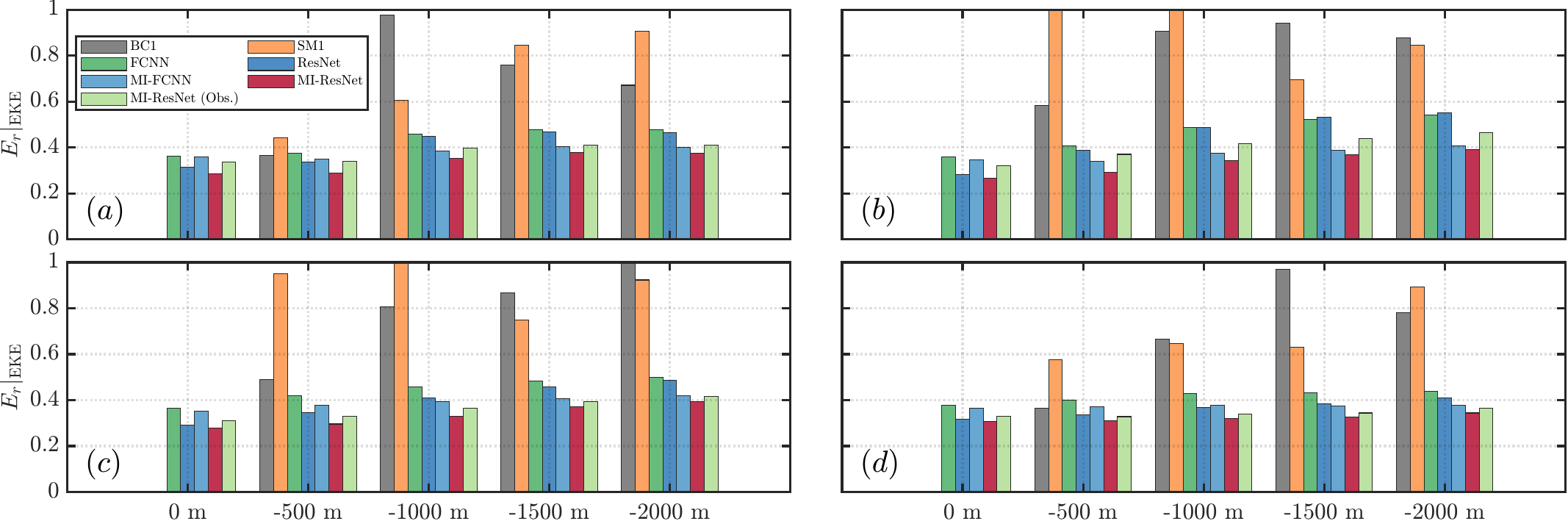}
   \caption{The temporally averaged relative error $E_{r}$ of $\mathrm{EKE}$ produced by different models (BC1, SM1, FCNN, ResNet, MI-FCNN, and MI-ResNet) for the test dataset (2017-2020): (a) GSR, (b) KR, (c) AR, (d) BMCR.}
   \label{Fig: Er_region_four}
 \end{center}
\end{figure}

In Figures~\ref{Fig: R2_region_four} and \ref{Fig: Er_region_four}, we perform a quantitative comparison of neural network models (FCNN, ResNet, MI-FCNN, and MI-ResNet) with traditional physics-based models (BC1 and SM1) to evaluate their performance in predicting the vertical structure of EKE. As shown in Figure~\ref{Fig: R2_region_four}, we assume that the surface EKE of BC1 and SM1 models uses the ground truth $\mathrm{EKE}_\mathrm{Truth}^{s}$, so the surface coefficients of efficiency $R^2$ of these two models are 1. The BC1 model performs well in predicting EKE at 500 m depth. The $R^2$ of BC1 and SM1 models in the Gulf Stream region are greater than 0.60, but we also note that $R^2$ varies in different regions, which is related to the ocean currents in different regions. As the depth increases, the performance of the BC1 and SM1 models decreases, causing the $R^2$ value to approach 0 or even become negative. This is because the correlation between the vertical structure of $\Phi^2(z)$ and the EKE in the deep ocean requires further improvement. The FCNN and ResNet models perform consistently well at sea surface with $R^2$ larger than 0.72 and 0.80, respectively, leveraging sufficient surface information for accurate predictions. Meanwhile, by considering the input tensor with multiple locations in the surface space, the $R^2$ predicted by the ResNet model is 0.08 higher than that predicted by the FCNN model. As shown in Table~\ref{Table: Network inputs}, the inputs of the FCNN and ResNet models only contain sea surface variables, which causes $R^2$ to decrease with the depth. In particular, in the Agulhas region, $R^2$ of the FCNN and ResNet models are less than 0.41 and 0.44 at 2000~m depth. 
Furthermore, the MI-FCNN and MI-ResNet models simultaneously incorporate both the sea surface and subsurface information, introducing the subsurface filtered velocity gradients closely related to EKE through the Taylor-series expansion of EKE. This makes the MI-FCNN and MI-ResNet models not only achieve high $R^2$ at the sea surface, but also perform well in the deep ocean. For example, The $R^2$ are greater than 0.74 and 0.81 at the sea surface, and exceed 0.58 and 0.62 at a depth of 2000 m. Besides, the MI-ResNet (Obs.) model with transfer learning performs well with the observational data, which achieves $R^2$ exceeding 0.78 at the surface and 0.54 at 2000~m for four high EKE regions (GSR, KR, AR, and BMCR). This shows that the MI-ResNet model has strong generalization.

Figure~\ref{Fig: Er_region_four} shows the relative errors $E_{r}$ between reconstructed $\mathrm{EKE}_\mathrm{Model}$ from different models and ground truth ($\mathrm{EKE}_\mathrm{Truth}$). These results are consistent with the correlation of efficiencies $R^2$ presented in Figure~\ref{Fig: R2_region_four}. $E_{r}$ of the BC1 and SM1 models are zero at the sea surface and increase with depth. The sea surface $E_{r}$ of the FCNN and ResNet models in four regions (GSR, KR, AR, and BMCR) are less than 0.38 and 0.32, respectively. However, at 2000~m depth, $E_r$ varies greatly in different regions, ranging from 0.40 to 0.55. This shows that using only sea surface data to predict deep ocean EKE in different regions will result in large fluctuations. Furthermore, by incorporating a subsurface input branch in the network, both MI-FCNN and MI-ResNet models exhibit decreasing relative errors $E_{r}$ from the sea surface to 2000 m depth. For example, in the Gulf Stream region, the MI-FCNN model predicts $E_{r}$ of 0.36 at the sea surface and 0.40 at 2000~m depth. The MI-ResNet model, on the other hand, predicts $E_{r}$ of 0.29 and 0.38 at the sea surface and 2000 m depth, respectively. The MI-ResNet model stands out as the best-performing model. The results of the MI-ResNet (Obs.) model in the four regions are consistent, with Er less than 0.34 and 0.47 at the surface and 2000 m depth, respectively.

\begin{figure}[tbh!]
 \begin{center}
   \includegraphics[width=0.45\linewidth]{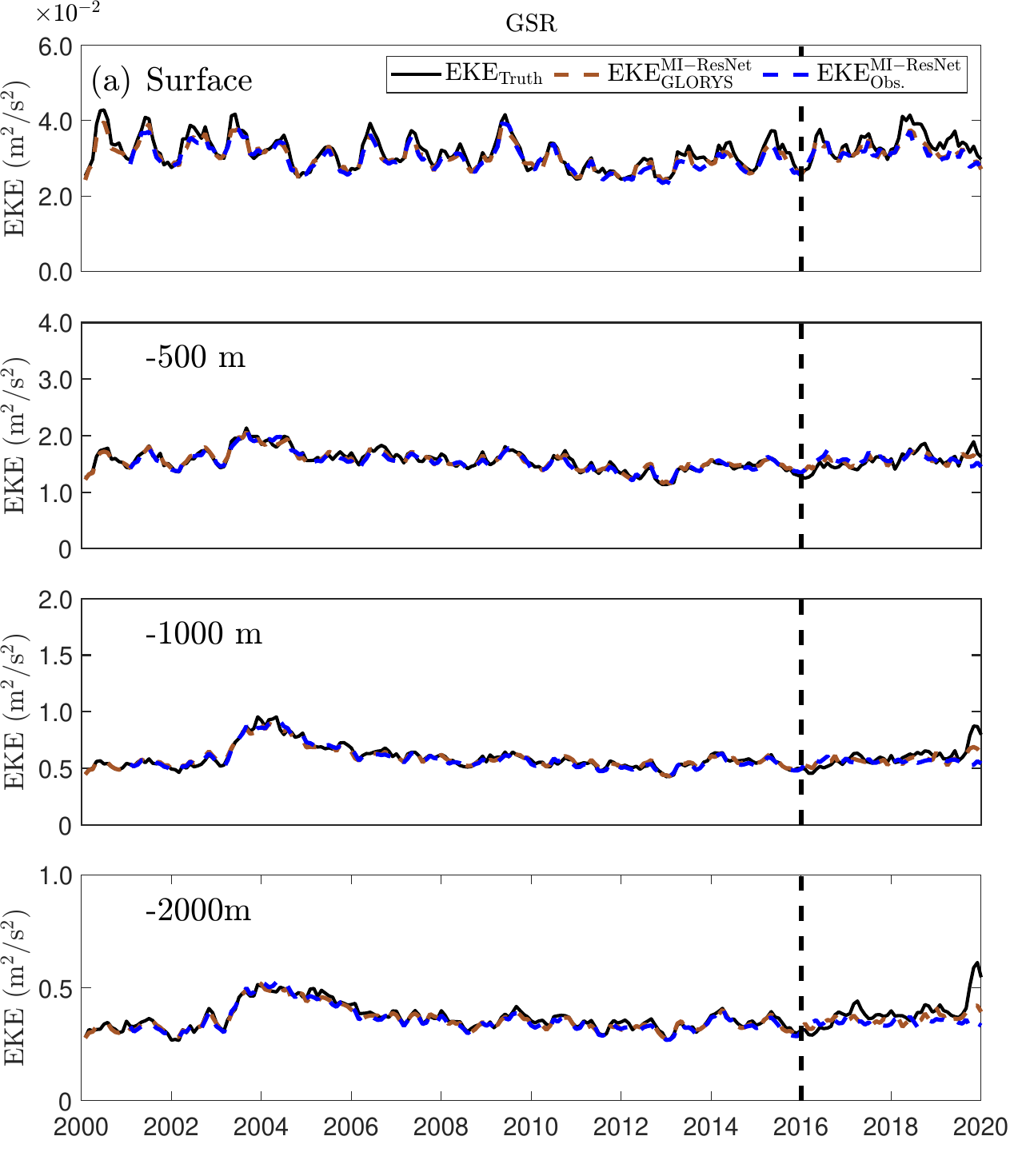}
   \includegraphics[width=0.45\linewidth]{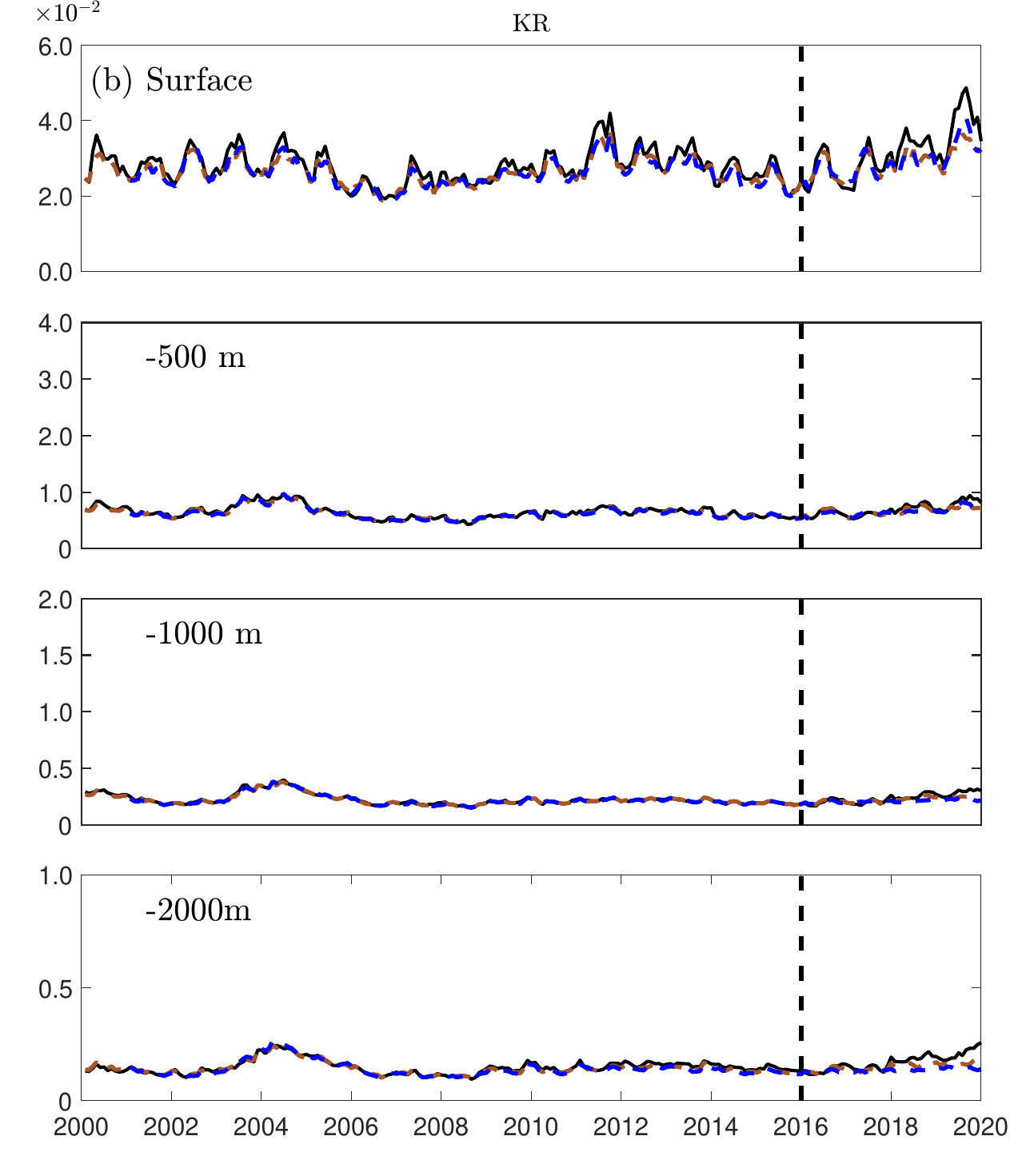}      
   \includegraphics[width=0.45\linewidth]{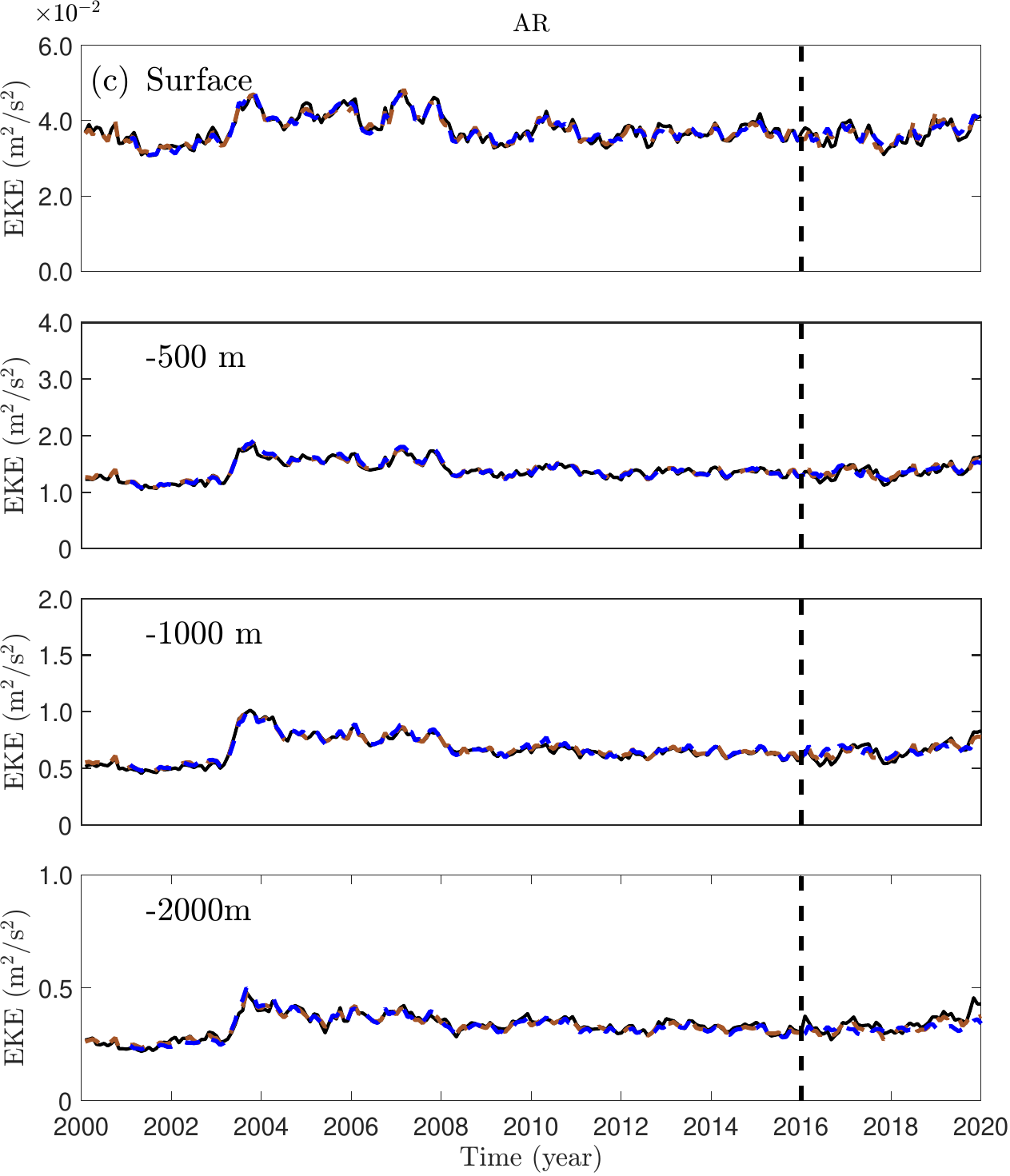}
   \includegraphics[width=0.45\linewidth]{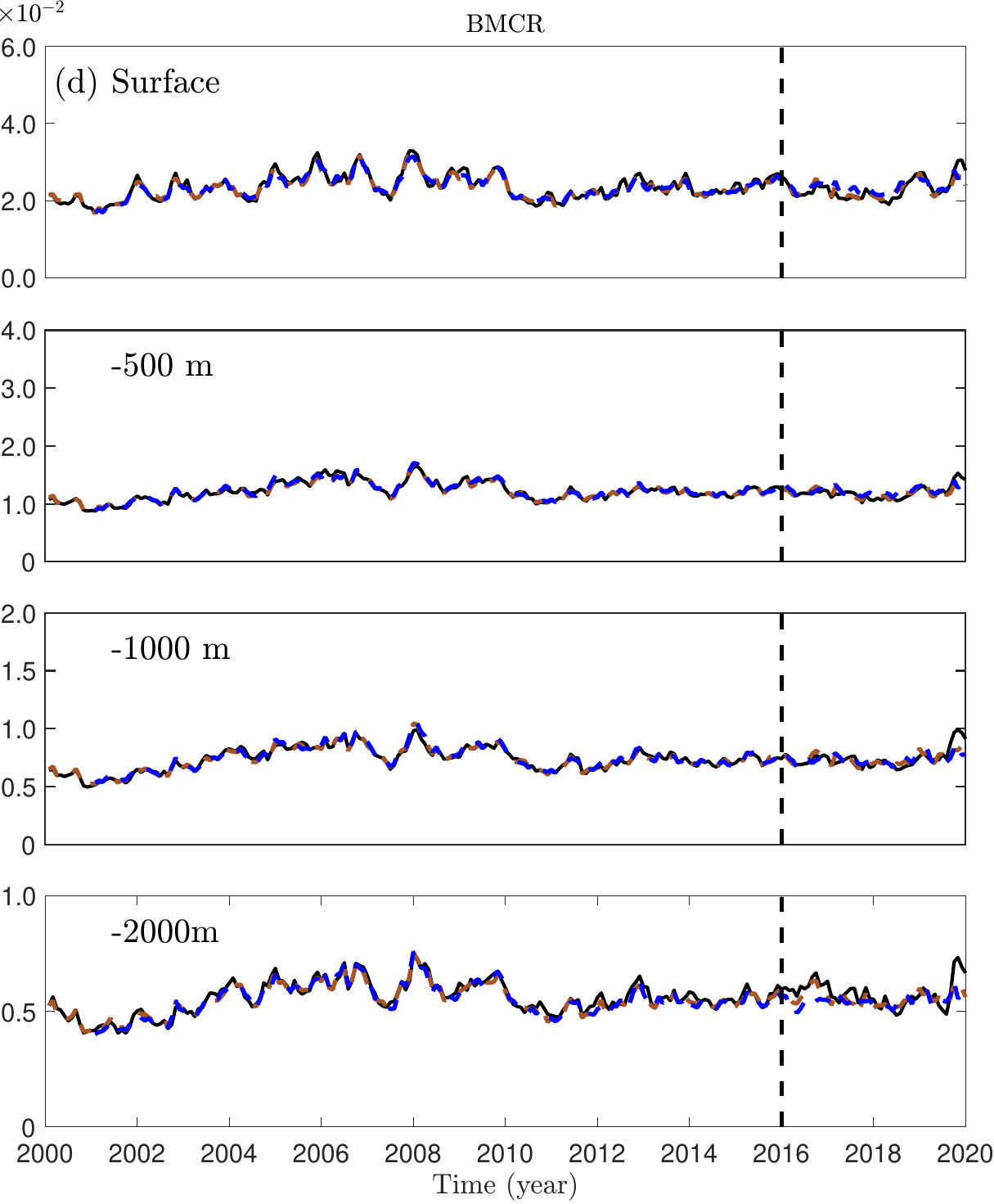}
   \caption{The time series of the regional average EKE diagnosed from the eddy-resolving GLORYS reanalysis data and reconstructed from the MI-ResNet model with both eddy-resolving GLORYS reanalysis and observational data: (a)~GSR, (b)~KR, (c)~AR, (d)~ BMCR. The dashed black lines divide data into training and test datasets.}
   \label{Fig: Timeseries_EKE_region}
 \end{center}
\end{figure}

\begin{figure}[tbh!]
 \begin{center}
   \includegraphics[width=0.45\linewidth]{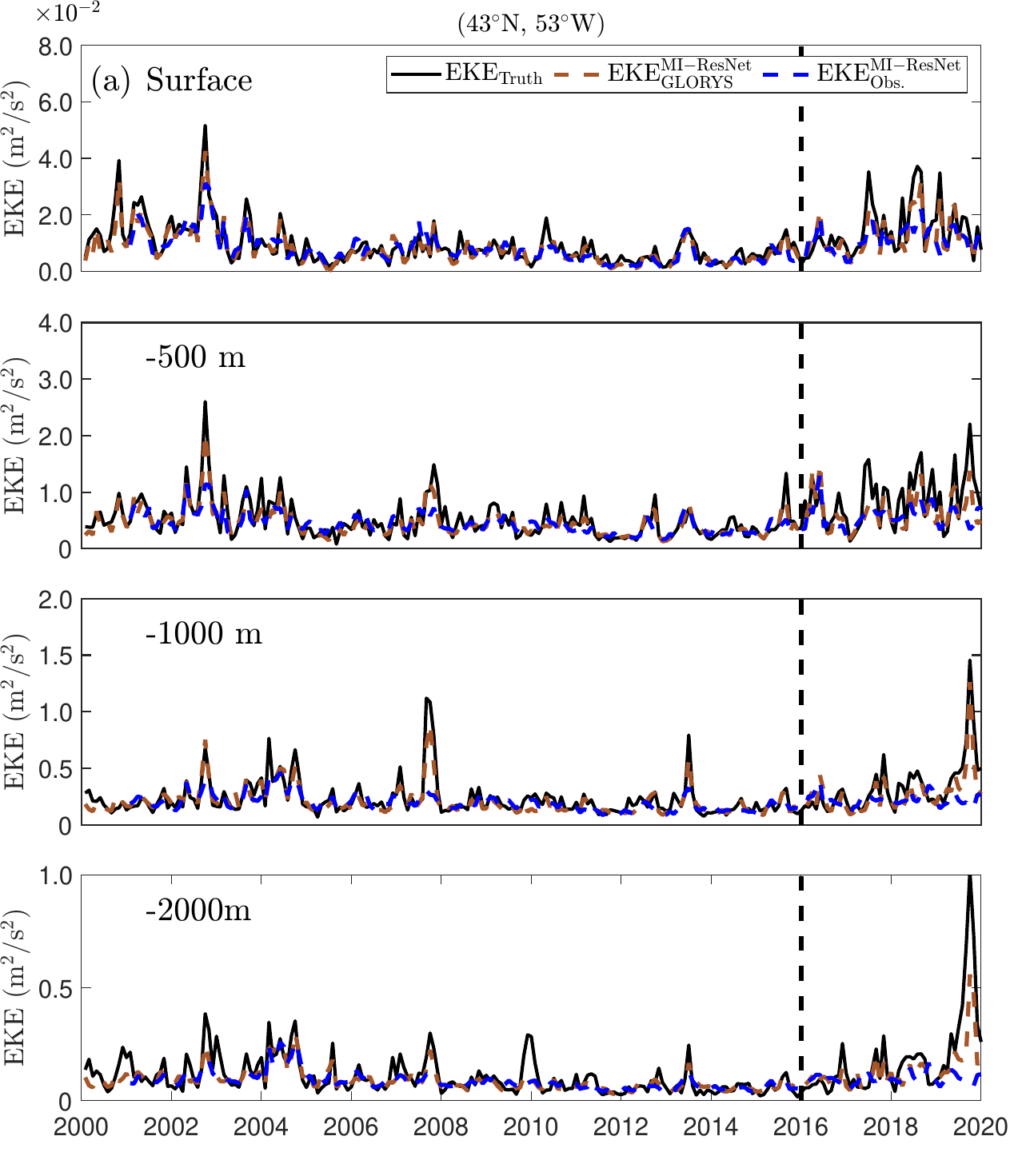}
   \includegraphics[width=0.45\linewidth]{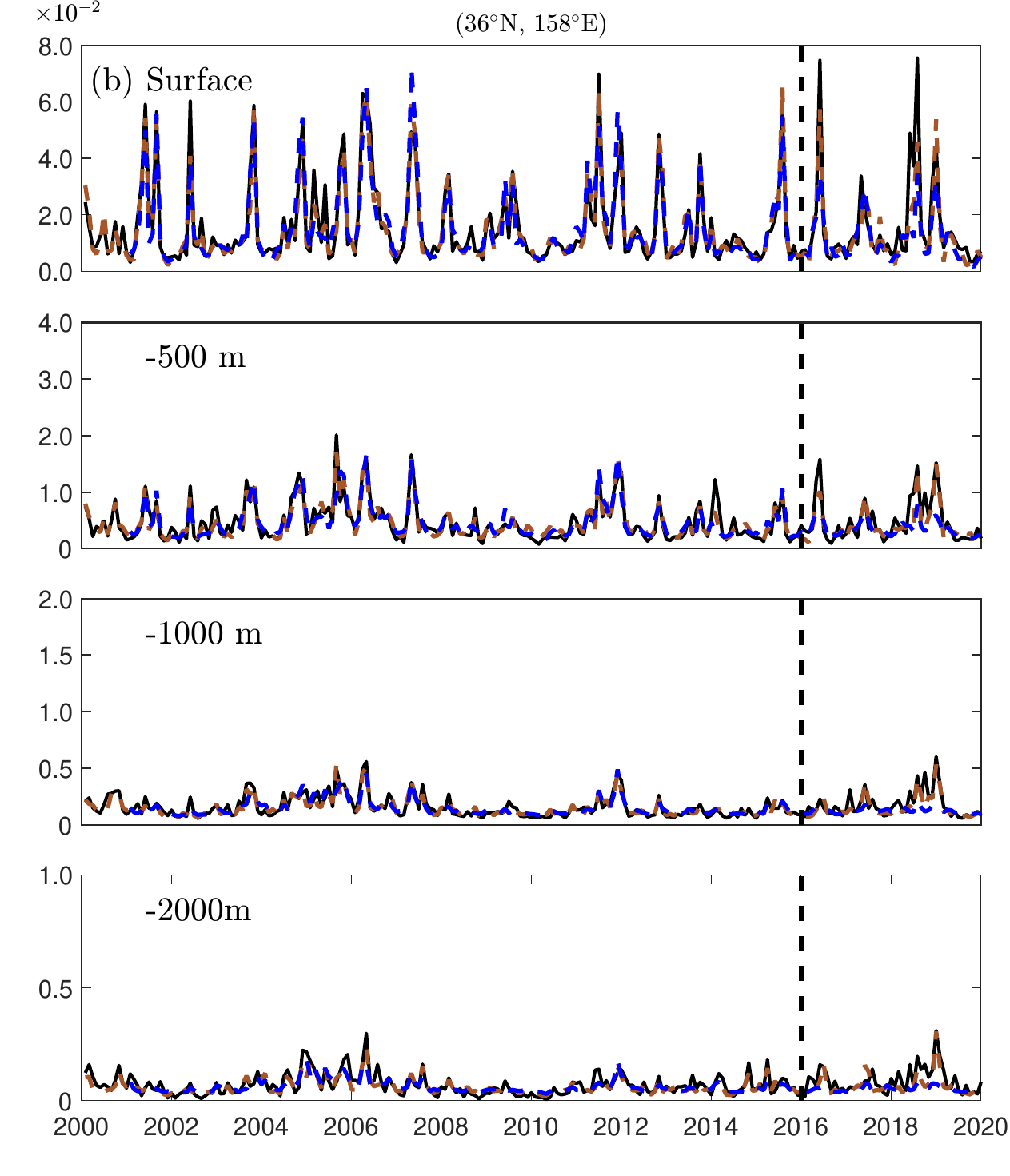}   
   \includegraphics[width=0.45\linewidth]{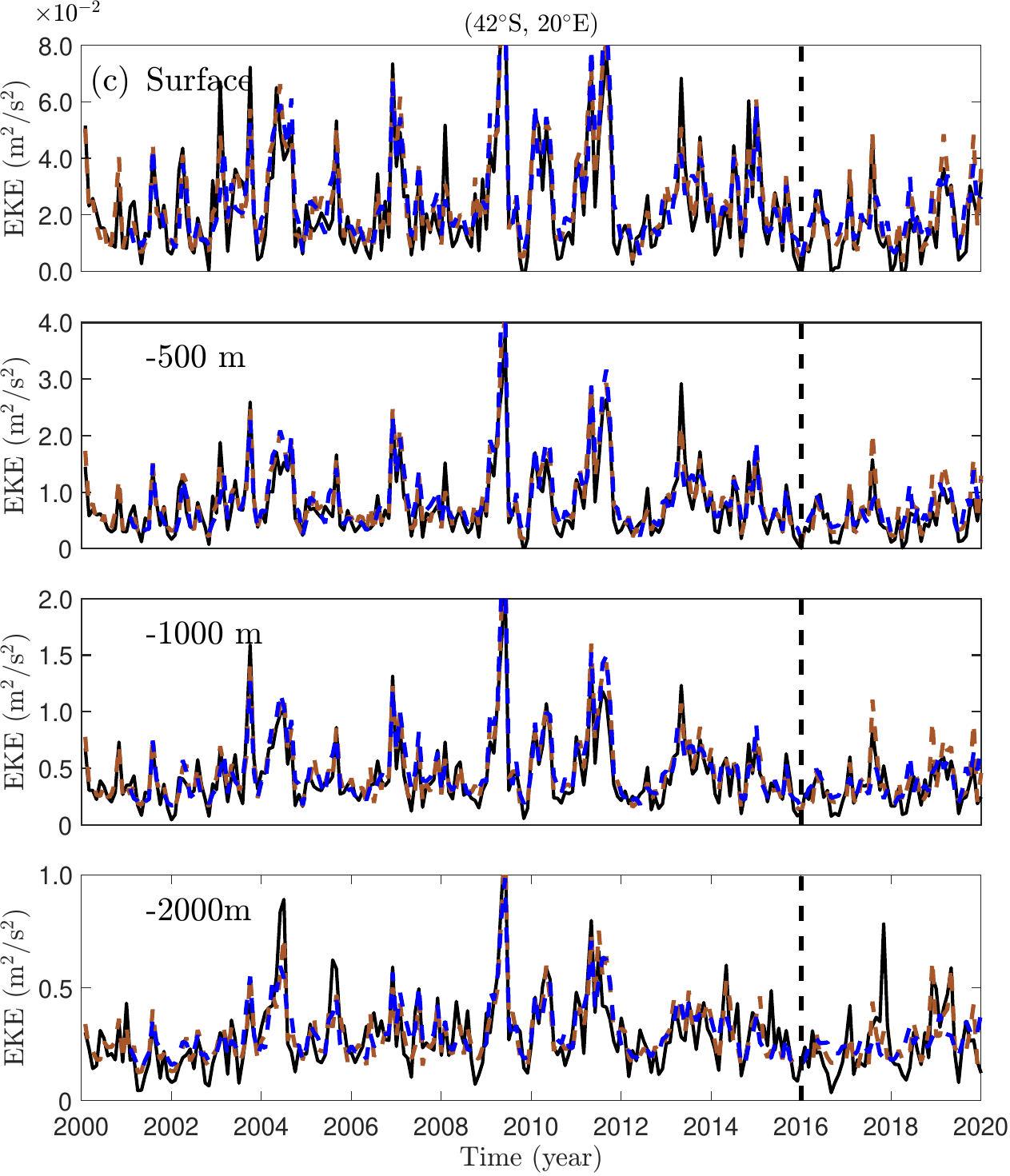}
   \includegraphics[width=0.45\linewidth]{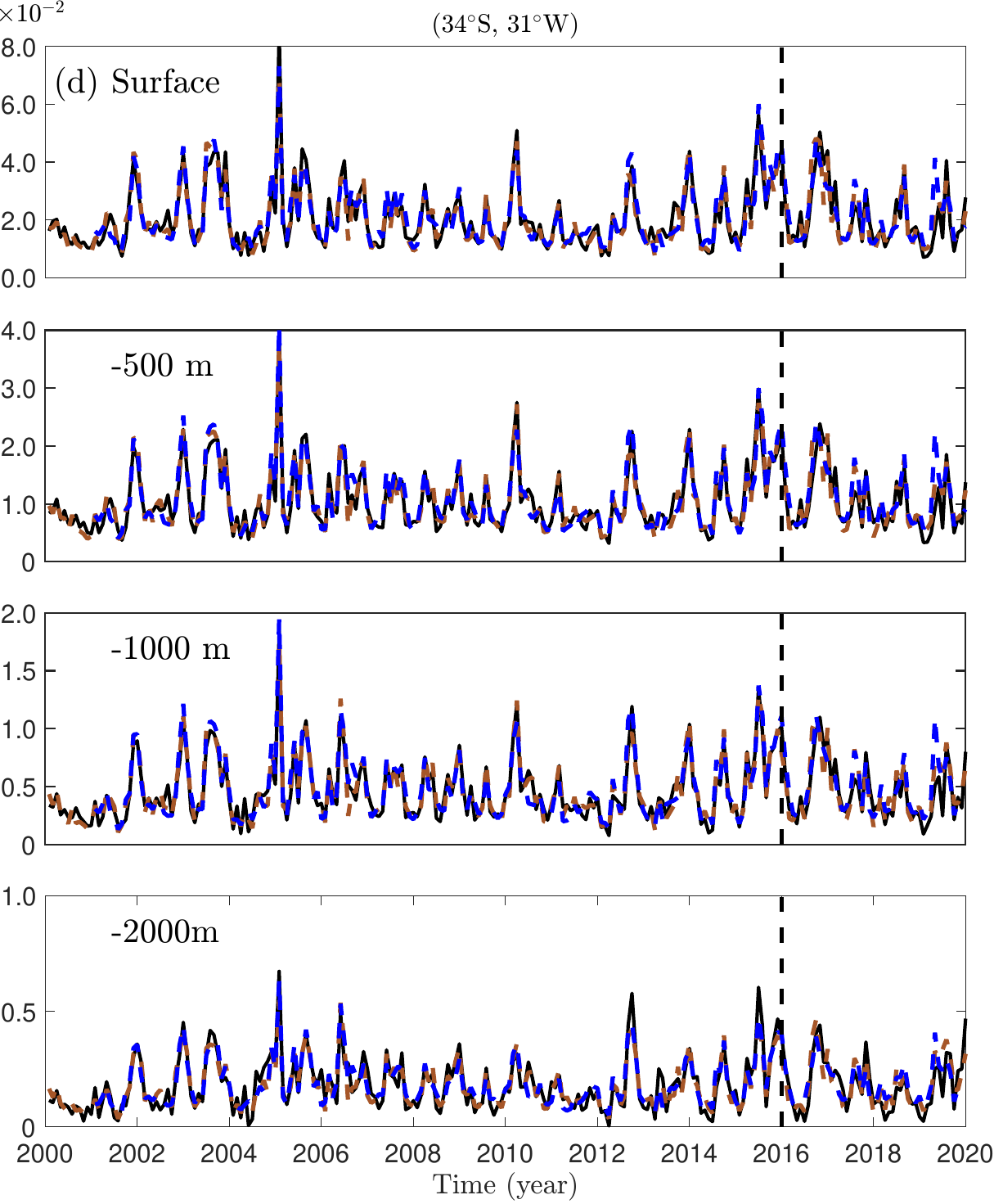}
   \caption{The time series of the EKE diagnosed from the eddy-resolving GLORYS reanalysis data and reconstructed from the MI-ResNet model with both eddy-resolving GLORYS reanalysis and observational data at different locations:
   (a)~GSR ($43^\circ\mathrm{N}$, $53^\circ\mathrm{W}$), (b)~KR ($36^\circ\mathrm{N}$, $158^\circ\mathrm{E}$), (c)~AR ($42^\circ\mathrm{S}$, $20^\circ\mathrm{E}$), (d)~ BMCR ($34^\circ\mathrm{S}$, $31^\circ\mathrm{W}$). The dashed black lines divide data into training and test datasets.}
   \label{Fig: Timeseries_EKE_point}
 \end{center}
\end{figure}

The time series of the regional average EKE reconstructed from the MI-ResNet model for four regions are shown in Figure~\ref{Fig: Timeseries_EKE_region}. The mean and variance of EKE decrease with increasing depth in the four regions. For example, in the Gulf Stream region, the mean value of $\mathrm{EKE}_\mathrm{Truth}$ decreases from 0.0317 at the sea surface to 0.0037 at a depth of 2000 m. Similarly, the variance of $\mathrm{EKE}_\mathrm{Truth}$ decreases from $1.77\times10^{-5}$ at the sea surface to $3.43\times10^{-7}$ at a depth of 2000 m. Surface EKE exhibits strong seasonal variability, gradually diminishing with increasing depth and correlating with surface driving forces. The MI-ResNet model captures these seasonal variations effectively and maintains consistent trends with ground truth in both training and test datasets. Consistent with the results presented in Figures~\ref{Fig: R2_region_four} and \ref{Fig: Er_region_four}, it can be observed that the prediction errors increase with depth. This phenomenon can be attributed to two main factors: first, the intensity of EKE decreases with increasing depth, accompanied by more randomness; second, the correlation between EKE in the deep ocean and surface information becomes weaker with increasing depth.

Figure~\ref{Fig: Timeseries_EKE_point} illustrates the time series of EKE at four spatial locations, which are located in four high EKE regions as shown in Figure~\ref{Fig: EKE_u}. The intensity of EKE at all locations and depths shows vigorous interannual variations between 2001 and 2020. At the specific location of ($42^\circ\mathrm{S}$, $20^\circ\mathrm{E}$), the predictions of EKE by the MI-ResNet model based on eddy-resolving GLORYS reanalysis and observational data are in good agreement with the patterns of the ground truth results. Benefiting from its effective integration of both surface and subsurface information, the MI-ResNet model maintains strong performance in ensuring consistency between eddy-resolving GLORYS reanalysis and observational data. Additionally, Equation~(\ref{Eqn: Fg}) guarantees the physical connection between EKE and available data, which is crucial for accurately predicting the dynamics of oceanic eddies.

\subsection{The region with weak eddies}\label{Sec: Region with weak eddies}

\begin{figure}[tbh!]
 \begin{center}
  \includegraphics[width=0.9\linewidth]{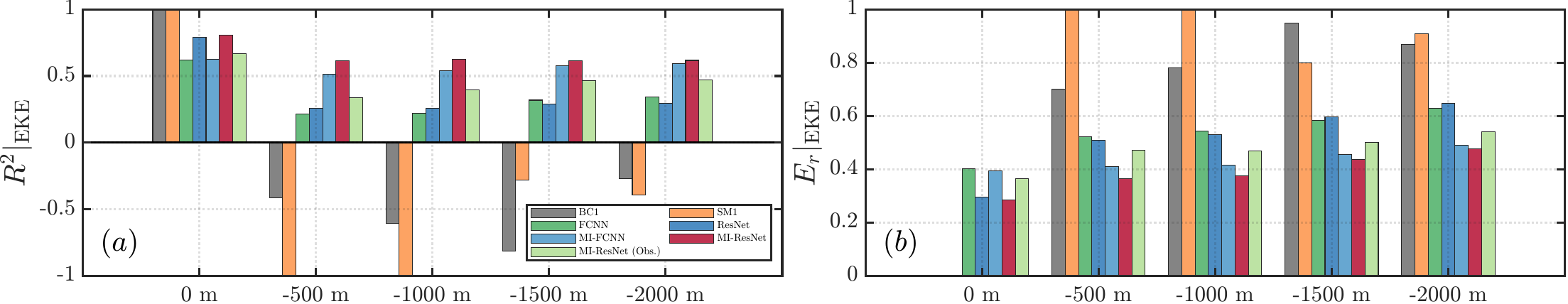}
   \caption{The temporally averaged coefficients of efficiency $R^2$ and relative error $E_{r}$ of $\mathrm{EKE}$ over the SPOR produced by different models (BC1, SM1, FCNN, ResNet, MI-FCNN, and MI-ResNet) for the test dataset (2017-2020): (a) $R^2$, (b) $E_{r}$.}
   \label{Fig: R2_Er_SPOR}
 \end{center}
\end{figure}

In the previous section, we discussed the performance of various models in four global hotspots of EKE. Apart from these high EKE regions, the EKE intensity in other global regions is weaker, often by an order of magnitude. Figure~\ref{Fig: R2_Er_SPOR} presents the coefficients of efficiency and relative errors of different models within the South Pacific Ocean region. In this weak EKE region, it is challenging to predict subsurface EKE using the BC1 and SM1 models. $R^2$ predicted by the FCNN and ResNet models are 0.62 and 0.79 at the surface, and 0.34 and 0.30 at 2000~m depth, respectively. For $E_{r}$, the FCNN and ResNet models predict 0.40 and 0.30 at the surface, and 0.63 and 0.65 at 2000~m depth, respectively. The input tensor of the sea surface variables significantly improves the performance of the ResNet model in surface EKE prediction. By introducing the subsurface input branch with filtered velocity gradients, the MI-FCNN and MIResNet models significantly improve the prediction of EKE in the deep ocean. For example, at 2000~m depth, the $R^2$ predicted by the MI-FCNN and MIResNet models are 0.60 and 0.62 respectively, while $E_{r}$ are 0.49 and 0.48. Leveraging the advantages of both the horizontal tensor of surface inputs and subsurface filtered velocity gradients, the MI-ResNet model stands out as the top-performing model.

\begin{figure}[tbh!]
 \begin{center}
   \includegraphics[width=0.45\linewidth]{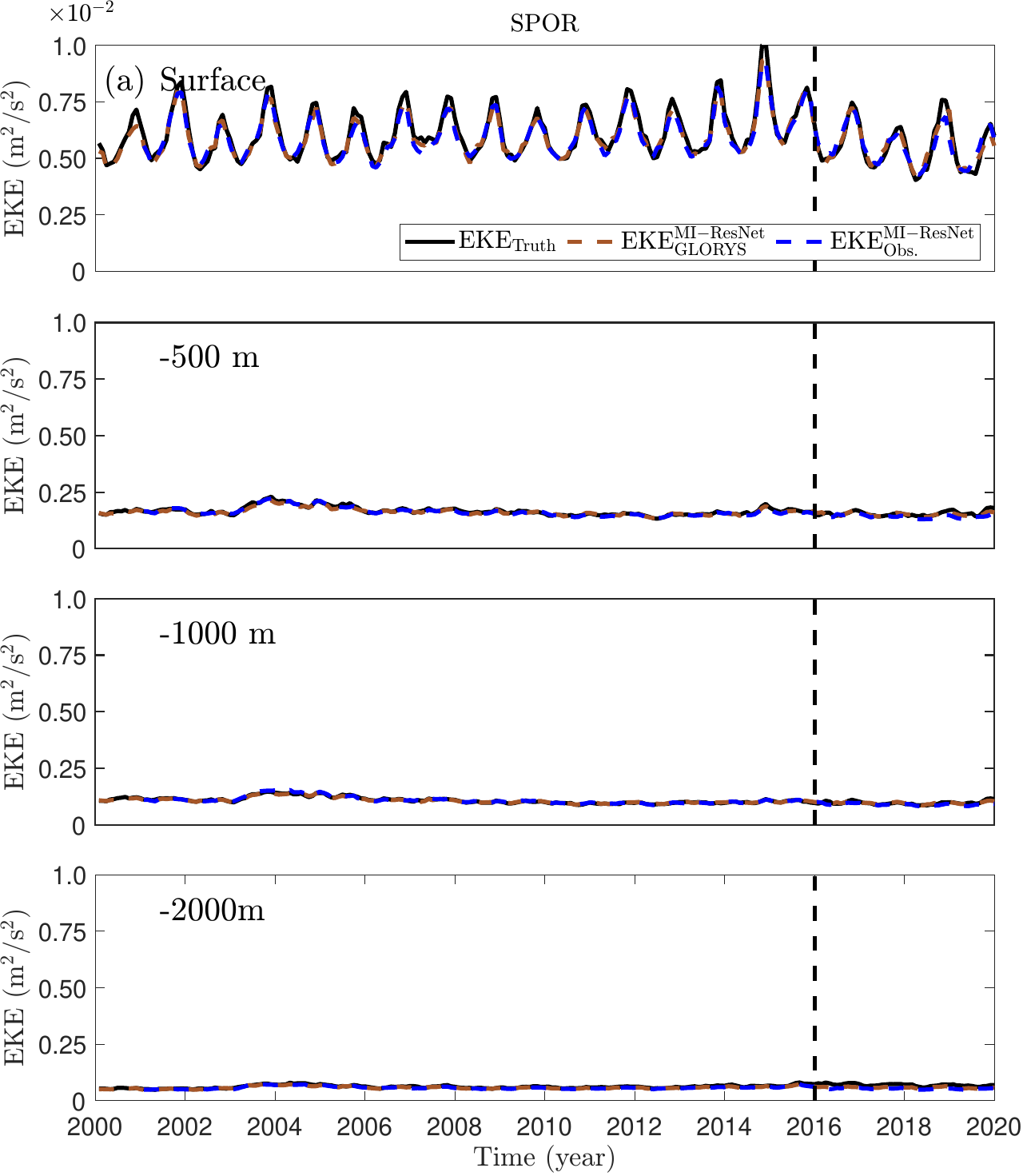}
   \includegraphics[width=0.45\linewidth]{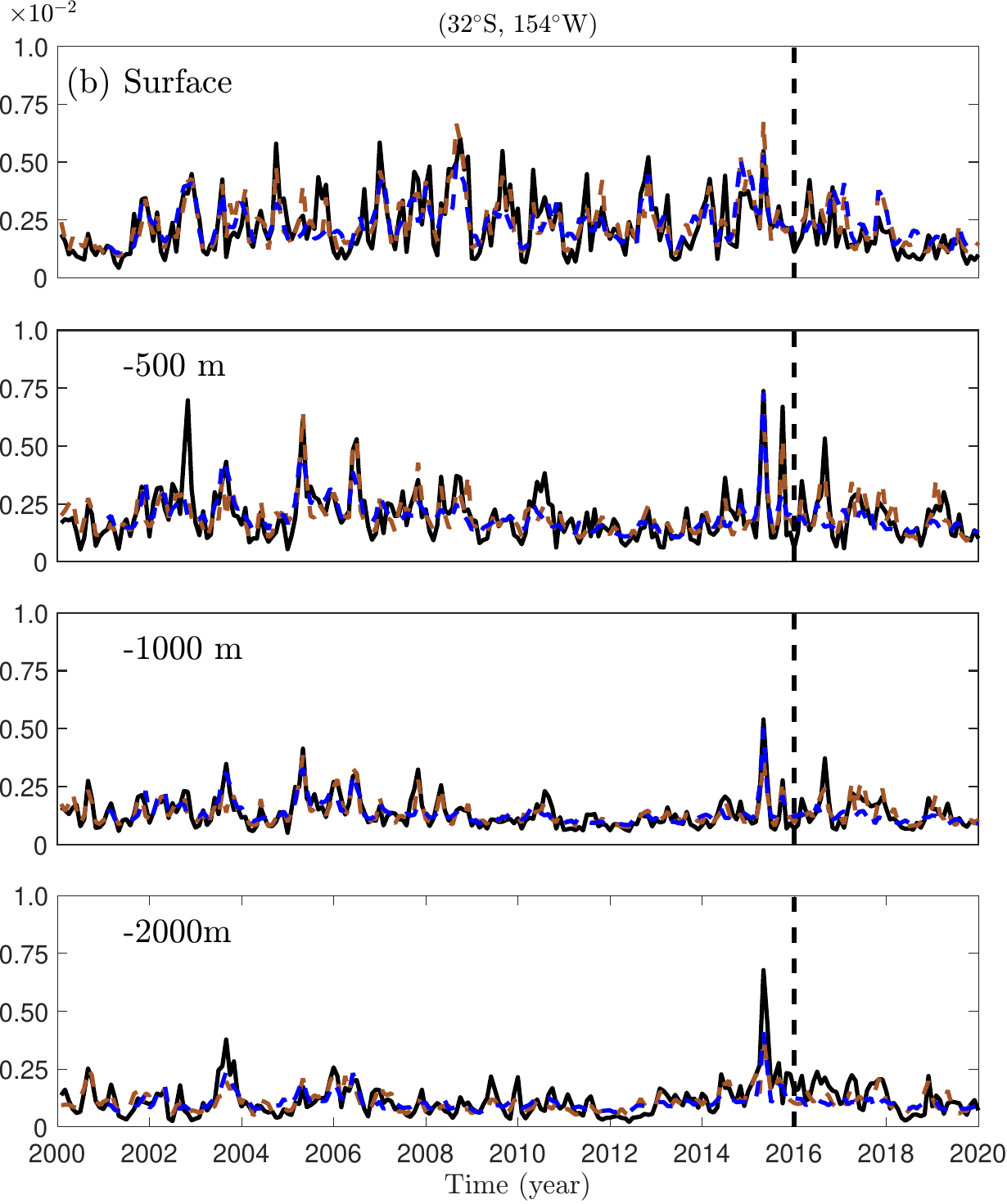}
   \caption{The time series of EKE diagnosed from the eddy-resolving GLORYS reanalysis data and reconstructed from the MI-ResNet model with both eddy-resolving GLORYS reanalysis and observational data at SPOR: (a)~region-weighted mean EKE, (b) Single location  ($32^\circ\mathrm{S}$, $154^\circ\mathrm{W}$). The dashed black lines divide data into training and test datasets.}
   \label{Fig: Timeseries_EKE_SPOR}
 \end{center}
\end{figure}

Figures~\ref{Fig: Timeseries_EKE_SPOR}a and b present the time series of EKE at the surface, 500 m, 1000 m, and 2000 m depths in the South Pacific Ocean. The intensity of surface EKE is an order of magnitude smaller than the results discussed in the previous section, and is highly periodic. The MI-ResNet model, based on eddy-resolving GLORYS reanalysis and observational data, can effectively capture the near-periodic structure of surface regional mean EKE from 2017 to 2020. As the depth increases, the MI-ResNet model can also capture the trend of EKE at ($32^\circ\mathrm{S}$, $154^\circ\mathrm{W}$).

These findings emphasize the importance of surface data and subsurface vertical profiles of variables in reconstructing the vertical distribution of EKE. Surface information mainly helps to reconstruct EKE near the surface, while incorporating subsurface geostrophic variables based on the thermal wind relation greatly enhances EKE reconstruction in deep ocean. This enhancement is reflected in the coefficient of efficiency and relative error involving deep ocean EKE prediction.

\subsection{Global analysis}\label{Sec: Global analysis}

As demonstrated in the Sections~\ref{Sec: Intense eddies regions} and \ref{Sec: Region with weak eddies}, the MI-ResNet model performs well in the training region. To further examine its generalization, we evaluate its performance in the global ocean. Satellite observations show that regions with high surface EKE are primarily concentrated along western boundary currents, where the flows are approximately in geostrophic balance and exhibit high kinetic energy. Meanwhile, regions with lower kinetic energy may experience comparable strengths in both geostrophic and ageostrophic motions \cite{Khatri-2021}. For the global ocean analysis, the equatorial region ($[10^\circ\mathrm{S}, 10^\circ\mathrm{N}]$) 
is omitted due to limitations in the geostrophic approximation, which is influenced by the Coriolis effect near the equator. Additionally, regions beyond $60^\circ$ latitude are excluded due to challenges in satellite spatial coverage~\cite{Martinez-Moreno-2022}. In order to more quantitatively examine the performance of the MI-ResNet model in the global ocean, we divide the global ocean into the following regions: the training region (TR), the non-training region (NONTR), the Southern Ocean ($[30^\circ\mathrm{S}, 60^\circ\mathrm{S}]$, SO), the Southern Hemisphere ($[10^\circ\mathrm{S}, 60^\circ\mathrm{S}]$, SH), the Northern Hemisphere ($[10^\circ\mathrm{N}, 60^\circ\mathrm{N}]$, NH), and global ocean ($[10^\circ\mathrm{S}, 60^\circ\mathrm{S}]$ and  $[10^\circ\mathrm{N}, 60^\circ\mathrm{N}]$, GO).

\begin{figure}[tbh!]
 \begin{center}
   \includegraphics[width=1.0\linewidth]{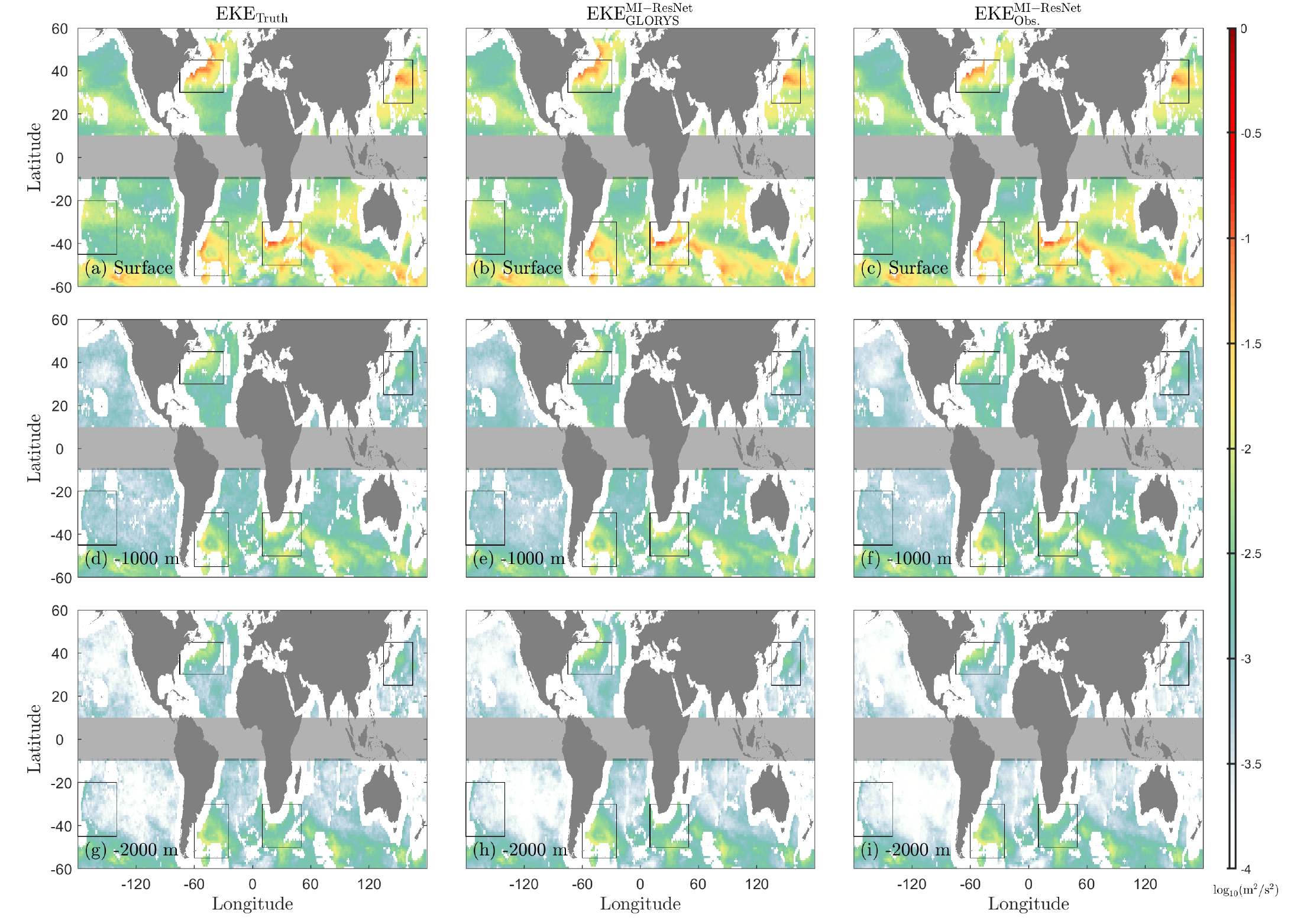}     
   \caption{The contours of temporal averaged EKE over 2017 to 2020 years at different depths around the global ocean: (a) Ground truth (0~m), (b) MI-ResNet (GLORYS, 0~m), (c) MI-ResNet (Observation, 0~m), (d) Ground truth (-1000~m), (e) MI-ResNet (GLORYS, -1000~m), (f) MI-ResNet (Observation, -1000~m), (g) Ground truth (-2000~m), (h) MI-ResNet (GLORYS, -2000~m), (i) MI-ResNet (Observation, -2000~m).}
   \label{Fig: EKE_global}
 \end{center}
\end{figure}

When applying the locally trained MI-ResNet model globally, we ensure full coverage of all longitudes by cyclically extending the input variables. This method ensures that the input tensor at $i=N_{x}$ includes information from $i=1$~\cite{Guillaumin-2021}. Figure~\ref{Fig: EKE_global} compares the EKE reconstructed by the MI-ResNet model based on eddy-resolving GLORYS reanalysis and observational data. The intensity of EKE decreases with increasing depth. Both $\mathrm{EKE}_\mathrm{GLORYS}^\mathrm{MI-ResNet}$ and $\mathrm{EKE}_\mathrm{Obs.}^\mathrm{MI-ResNet}$ align closely with the $\mathrm{EKE}_\mathrm{Truth}$ at different depths. The reconstructed EKE is significantly stronger in the western boundary currents and is approximately an order of magnitude lower in the South Pacific region, which is consistent with $\mathrm{EKE}_\mathrm{Truth}$. The MI-ResNet's prediction of EKE based on global ocean further validates the robustness and reliability of the model.


\begin{figure}[tbh!]
 \begin{center}
  \includegraphics[width=0.9\linewidth]{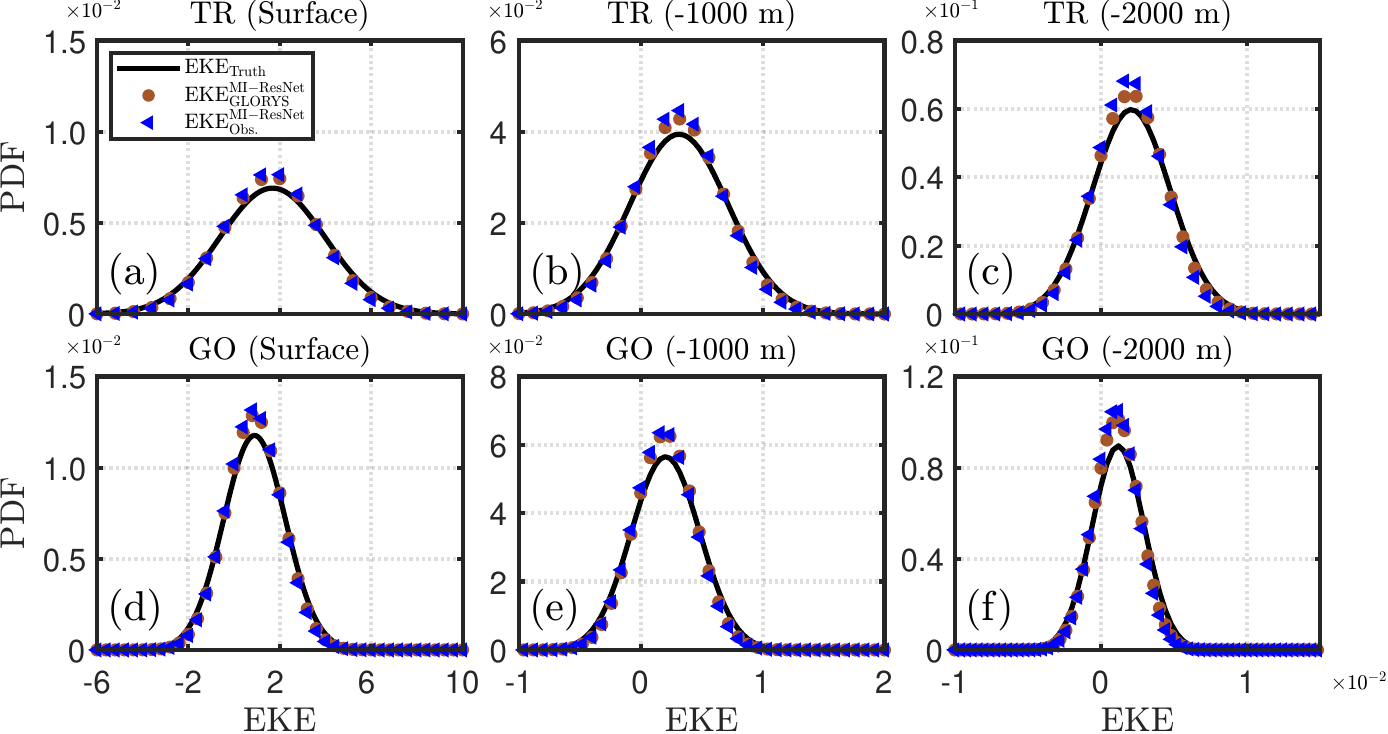}
   \caption{The probability density functions of $\mathrm{EKE}_\mathrm{Truth}$, $\mathrm{EKE}_\mathrm{GLORYS}^\mathrm{ResNet}$, and $\mathrm{EKE}_\mathrm{Obs.}^\mathrm{ResNet}$ in the training region (TR) and global ocean (GO): (a) TR: surface, (b) TR: -1000 m, (c) TR: -2000 m, (d) GO: surface, (e) GO: -1000 m, (f) GO: -2000 m.}
   \label{Fig: PDF_global}
 \end{center}
\end{figure}

\begin{figure}[tbh!]
 \begin{center}
  \includegraphics[width=0.9\linewidth]{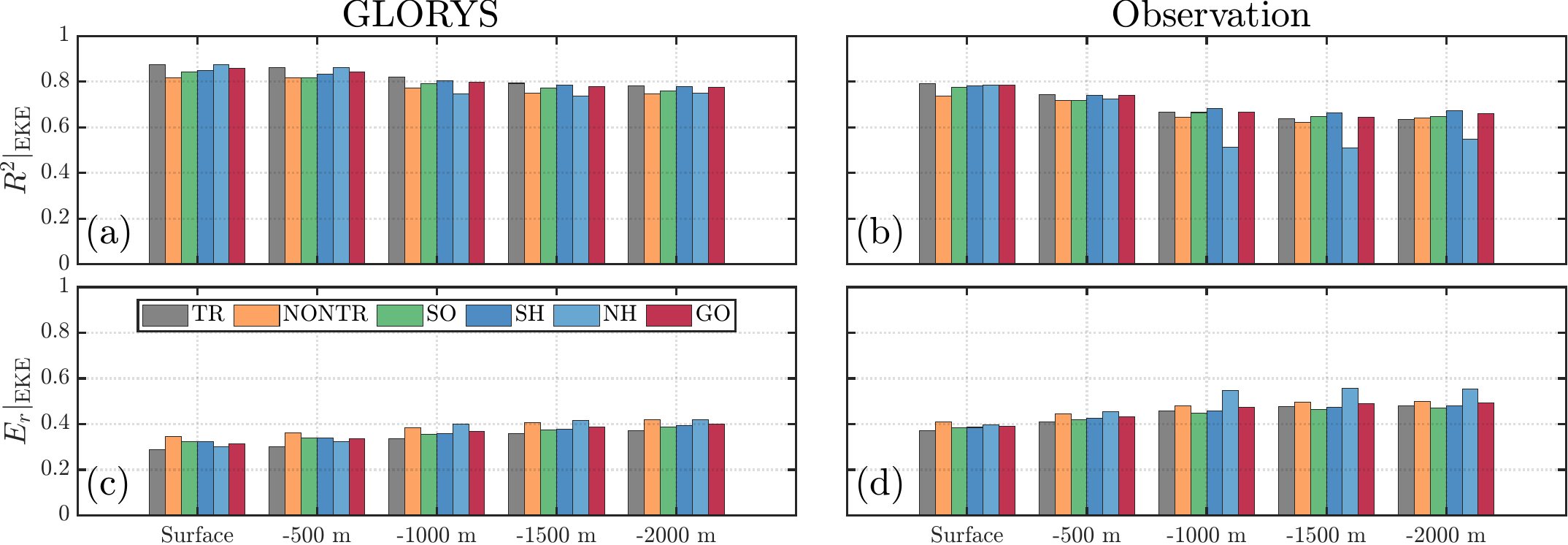}
   \caption{The temporally averaged coefficients of efficiency $R^2$ and relative error $E_{r}$ of $\mathrm{EKE}$ over the global ocean produced by the MI-ResNet model based on the eddy-resolving GLORYS reanalysis and observational data: (a) $R^2$ (MI-ResNet: GLORYS), (b) $R^2$ (MI-ResNet: Observation), (c) $E_{r}$ (MI-ResNet: GLORYS), (d) $E_{r}$ (MI-ResNet: Observation). Here, `TR', `NONTR', `SO', `SH', `NH', and `GO' are abbreviations for Training regions, Non-training regions, southern ocean, southern hemisphere, northern hemisphere, and global ocean, respectively.}
   \label{Fig: R2_Er_global}
 \end{center}
\end{figure}

Figure~\ref{Fig: PDF_global} shows the probability density functions (PDFs) of EKE for different depths in the training region and global ocean. Comparing the PDFs of the training region and global ocean results shows that the training data contains more high EKE regions. As depth increases, the intensity of EKE decreases. The PDFs of EKE predicted by the MI-ResNet model (brown and blue dots) are similar in shape and peak position to the ground truth results (solid black line), indicating that the MI-ResNet model performs effectively in these conditions. Figure~\ref{Fig: R2_Er_global}a quantifies the performance of the MI-ResNet model in predicting global EKE using the reanalysis input data. Across all regions, the MI-ResNet model performs well, consistently achieving $R^2$ exceeding 0.81 and $E_{r}$ below 0.35 for surface EKE prediction, while at 2000 m depth, it achieves $R^2$ over 0.74 and maintains $E_{r}$ under 0.42. Particularly for the global ocean, the MI-ResNet model achieves $R^2=0.859$ and $E_{r}=0.315$ at the surface, and $R^2=0.775$ and $E_{r}=0.399$ at 2000 m depth.

Furthermore, the generalization of the MI-ResNet model is evaluated with observational input data, as shown in Figure~\ref{Fig: R2_Er_global}b. The model achieves $R^2$ of 0.790, 0.738, 0.777, 0.782, 0.785, and 0.785 and maintains $E_{r}$ of 0.371, 0.409, 0.384, 0.386, 0.396, and 0.389 at the sea surface for the Training Region (TR), Non-Training Region (NONTR), Southern Ocean (SO), Southern Hemisphere (SH), Northern Hemisphere (NH), and Global Ocean (GO), respectively. At 2000 m depth in the global ocean, the model obtains $R^{2}=0.660$ and $E_{r}=0.491$, respectively. These results highlight the MI-ResNet model's robust performance across various regions and depths.

\section{Discussion and Conclusion}
\label{Sec: Discussion and conclusion}

This study introduces a multiple-input neural network framework (MI-FCNN and MI-ResNet) to reconstruct subsurface EKE with spatial filter from the surface to 2000 m depth. The MI-FCNN and MI-ResNet models, which are designed with more sophisticated architectures, effectively inherit the excellent prediction performance of the surface-input neural network models (FCNN and ResNet). Specifically, by integrating the subsurface input branch, the MI-FCNN and MI-ResNet models not only maintain but also enhance the strengths of the FCNN and ResNet models in predicting deep ocean EKE from surface inputs. The subsurface branch bridges the EKE with the subsurface filtered velocity gradients by utilizing the Taylor-series expansion of EKE, as outlined in Equation~\ref{Eqn: Fg}. In addition, the FCNN, ResNet, MI-FCNN and MI-ResNet models are trained in five representative regions of the ocean with the reanalysis data, as shown in Table~\ref{Table: Training region}. Compared to the BC1 and SM1 models, the neural network models demonstrate superior performance in reconstructing subsurface EKE. The ResNet model outperforms the FCNN model in predicting sea surface EKE because its surface input branch incorporates the horizontal spatial tensor, which can catch more spatial features of the surface variables. Both the MI-FCNN and MI-ResNet models perform effectively in the deep ocean, especially at 2000~m depth, by leveraging the strengths of both surface and subsurface input branches. Furthermore, the strong performance of the MI-ResNet model on reanalysis data can be generalized to observation data through transfer learning. The ResNet model exhibits robust performance globally with high coefficients of efficiency, offering promising potential for advancing ocean observations and the complex interactions of eddies within the global ocean system.

This study greatly benefits from the integration of machine learning and oceanography. It focuses on bridging global subsurface EKE with available observational information using machine learning techniques. Our study demonstrates that the subsurface EKE reconstruction is effective by integrating both surface and sparse subsurface variables into the multiple-input neural network framework. To explore the broader potential of the multiple-input neural network framework, it would be valuable to investigate whether our approach can be generalized to predict other oceanic variables, such as subsurface temperature, salinity, currents, subgrid-scale stress, kinetic energy flux, and biochemical tracers. This expansion is expected to provide further insights into vertical mixing and nutrient transport. Additionally, incorporating more physical measurements, such as chlorophyll maps that track the movement of plankton affected by ocean currents, could improve forecast accuracy and offer new insights. 

Spatial filtering distinguishes eddies based on spatial scales, while Reynolds averaging decomposes physical properties into ensemble-averaged components and fluctuations. Reynolds averaging has been widely employed in parameterization of ocean mesoscale eddies in ocean models. Our findings on predicting spatially filtered EKE by incorporating subsurface variables can also provide valuable insights into subsurface EKE using Reynolds averaging. Understanding how temporal averaged variables at different time scales (seasonal, interannual, and decadal) affect overall ocean circulation and climate patterns is critical to ocean modeling. Future studies will focus on quantifying the relationship between physical quantities at different time scales and unknown oceanic variables.

The integration of eddy-resolving GLORYS reanalysis and observational data helps alleviate the limitations of a single data source and evaluates the generalization of the neural network models. High-resolution SSH and SST serve as valuable inputs for estimating subsurface variables. The model proposed in this study can also be applied to ongoing research on accurately reconstructing gridded SSH and SST from satellite altimetry data. Combining SSH data from satellite altimetry with other datasets, such as Argo floats and ship-based measurements, provides a comprehensive understanding of the ocean. Future research will explore strategies that directly training machine learning models from observations. 

The advancements achieved in this study provide new insights for enhancing the prediction of unknown oceanic variables. Future research can extend these advancements by exploring the interactions among eddies at various scales, from small-scale turbulence to larger mesoscale eddies, to improve predictions of ocean dynamics. In addition, exploring new neural network architectures better suited for multi-input ocean data is another important step. For example, employing attention models that more effectively capture complex features of the input and Long Short-Term Memory (LSTM) models that account for time-series forecasting are promising directions. Overall, this research will serve as a valuable foundation for future studies of eddy dynamics, including scale interactions, parameterizations, and the impacts of mesoscale eddies on climate and marine ecosystems.

\section*{Open Research}
The Global Ocean Reanalysis 12 version 1 (GLORYS12V1) of the NEMO dataset used here is “\url{GLOBAL_MULTIYEAR_PHY_001_030}” (\url{https://doi.org/10.48670/moi-00021}). The satellite altimetry products from AVISO are produced by Ssalto/Duacs and distributed by EU Copernicus Marine and Environment Monitoring Service and can be found at \url{https://resources.marine.copernicus.eu/product-detail/SEALEVEL_GLO_PHY_L4_MY_008_047}. The SST data is obtained from the daily and 0.25$^\circ$-gridded Optimum Interpolation Sea Surface Temperature (OISST) v.2.1 (\url{https://www.ncei.noaa.gov/products/optimum-interpolation-sst}). The monthly gridded fields of temperature and salinity are produced by the In Situ Analysis System (ISAS20) (\url{https://www.seanoe.org/data/00412/52367/}).
The main codes used in this manuscript have been published at \url{https://doi.org/10.5281/zenodo.14328633}~\cite{ML_dataset}.

\acknowledgments
The authors thank Yan Wang, Huaiyu Wei, and Wenda Zhang for helpful discussions. This work is supported by the USTC Startup Program (KY2090000141), the National Natural Science Foundation of China (Grant Nos. 12388101), the Research Grants Council of Hong Kong (ECS26307720 and GRF16305321) and the Center for Ocean Research (CORE), a joint research center between QNLM and HKUST. 

\section*{Declaration of generative AI and AI‑assisted technologies in the writing process}
During the preparation of this work, the authors used ChatGpt in order to improve the manuscript grammatically. After using this tool/service, the authors reviewed and edited the content as needed and take full responsibility for the content of the publication.


%
%

\bibliography{Manuscript}

%
%
%
%

\end{document}


%
%


\title{Supporting Information for ``Impact of Parameterized Isopycnal Diffusivity on Shelf-Ocean Exchanges under Upwelling-Favorable Winds: Offline Tracer Simulations Augmented by Artificial Neural Network''}
%
%

%
%



 \authors{Chenyue Xie\affil{1} and Huaiyu Wei\affil{1} and Yan Wang\affil{1,2}}


\affiliation{1}{Department of Ocean Science, The Hong Kong University of Science and Technology, Hong Kong, China}
\affiliation{2}{Center for Ocean Research in Hong Kong and Macau, The Hong Kong University of Science and Technology, Hong Kong, China}

%
%

%

\begin{article}

%
%

\noindent\textbf{Contents of this file}
\begin{enumerate}
\item Caption for Movie S1
\item Caption for Movie S2
\item Caption for Movie S3

\end{enumerate}
\noindent\textbf{Additional Supporting Information (Files uploaded separately)}
\begin{enumerate}
\item Movie S1
\item Movie S2
\item Movie S3
\end{enumerate}

\newpage
\noindent\textbf{Introduction}

This Supporting Information document provides the Captions for Movie S1, Movie S2, and Movie S3.\\[10pt]

\noindent\textbf{Movie S1.} Year-long evolution of tracer contents of $C_1$ as functions of depth and offshore distance simulated by (a) the ensemble MITgcm simulations and the MAMEBUS runs with the isopycnal eddy diffusivity prescribed as (b) $\mathcal{K}_\mathrm{const}$, (c) $\mathcal{K}_\mathrm{MLT}$, (d) $\mathcal{K}_\mathrm{SMLT}$, (e) $\mathcal{K}_\mathrm{ASMLT}$, (f) $\mathcal{K}_\mathrm{FULL}$, (g) $\mathcal{K}_\mathrm{FULL}^\mathrm{ANN}$, (h) $\mathcal{K}_\mathrm{Redi}^\mathrm{ANN}$, and (i) $\kappa_\mathrm{Redi}$.
\vspace{30pt}

\noindent\textbf{Movie S2.} The isopycnal eddy diffusivity (a) diagnosed from MITgcm by \citeA{Wei-2021}, (b) parameterzed as $\mathcal{K}_\mathrm{FULL}^\mathrm{ANN}$ in the secondary reference simulation designed for fluxing $C_1$, updated on a daily basis for one model year, and (c) parameterzed as $\mathcal{K}_\mathrm{Redi}^\mathrm{ANN}$ in the secondary reference simulation designed for fluxing $C_1$, updated on a daily basis for one model year. Solid gray contours indicate selected time- and zonal-mean alongshore velocity in panel (a), and ensemble- and zonal-mean alongshore velocity in panels (b)--(c) (interval: 0.1 m$/$s; velocity of 0 m$/$s is plotted using bold contours). Dashed black contours indicate selected time- and zonal-mean potential temperature in panel (a),  and ensemble- and zonal-mean potential temperature in panels (b)--(c) (intervals: 1 $^{\circ}$C). 
\vspace{30pt}

\noindent\textbf{Movie S3.} The isopycnal eddy diffusivity (a) diagnosed from MITgcm by \citeA{Wei-2021}, (b) parameterzed as $\mathcal{K}_\mathrm{FULL}^\mathrm{ANN}$ in the secondary reference simulation designed for fluxing $C_2$ and $C_3$, animated every five model days for five model years, and (c) parameterzed as $\mathcal{K}_\mathrm{Redi}^\mathrm{ANN}$ in the secondary reference simulation designed for fluxing $C_2$ and $C_3$, animated every five model days for five model years. Solid gray contours indicate selected time- and zonal-mean alongshore velocity in panel (a), and ensemble- and zonal-mean alongshore velocity in panels (b)--(c) (interval: 0.1 m$/$s; velocity of 0 m$/$s is plotted using bold contours). Dashed black contours indicate selected time- and zonal-mean potential temperature in panel (a),  and ensemble- and zonal-mean potential temperature in panels (b)--(c) (intervals: 1 $^{\circ}$C). 

\end{article}

\bibliography{Manuscript}